\documentclass[useAMS,usenatbib]{mn2e}

\usepackage{graphics}
\usepackage{graphicx,subfigure}
\title[PSF interpolation]
{On Point Spread Function modelling: towards optimal interpolation}
\author[Joel Berg\'e et al]{Joel Berg\'e$^{1,2,3}$\thanks{E-mail:
jberge@phys.ethz.ch}, Sedona Price$^3$\thanks{Current address: Department of Astronomy, UC Berkeley, Berkeley, CA 94720, USA}, Adam Amara$^{1}$ and Jason Rhodes$^{2,3}$ \\
$^{1}$Department of Physics, ETH Zurich, Wolfgang-Pauli-Strasse 27, CH-8093 Zurich, Switzerland \\
$^2$Jet Propulsion Laboratory, California Institute of Technology, 4800 Oak Grove Drive, MS 169-327, Pasadena, CA 
91109, USA\\
$^3$California Institute of Technology, 1200 East California Blvd, Pasadena, CA 91125, USA}
\begin{document}

\date{Accepted ???. Received ???; in original form ???}

\pagerange{\pageref{firstpage}--\pageref{lastpage}} \pubyear{???}

\maketitle

\label{firstpage}

\begin{abstract}
Point Spread Function (PSF) modeling is a central part of any astronomy data analysis relying on measuring the shapes of objects. It is especially crucial for weak gravitational lensing, in order to beat down systematics and allow one to reach the full potential of weak lensing in measuring dark energy. A PSF modeling pipeline is made of two main steps: the first one is to assess its shape on stars, and the second is to interpolate it at any desired position (usually galaxies). We focus on the second part, and compare different interpolation schemes, including polynomial interpolation, radial basis functions, Delaunay triangulation and Kriging. For that purpose, we develop simulations of PSF fields, in which stars are built from a set of basis functions defined from a Principal Components Analysis of a real ground-based image. We find that Kriging gives the most reliable interpolation, significantly better than the traditionally used polynomial interpolation. We also note that although a Kriging interpolation on individual images is enough to control systematics at the level necessary for current weak lensing surveys, more elaborate techniques will have to be developed to reach future ambitious surveys' requirements.
\end{abstract}

\begin{keywords}
gravitational lensing - methods: data analysis - methods: statistical
\end{keywords}

\section{Introduction}

An accurate model of the Point Spread Function (PSF) is a crucial step in astronomical analyses relying on the estimation of galaxy shapes. For instance, an imperfect PSF model has been identified as a prominent systematic effect in weak gravitational lensing measurements (e.g. \citealt{step1,step2}), and much effort is underway to optimally model the PSF (e.g. \citealt{great10}). 
\cite{jarvis04} and \cite{jee07} have developed techniques based on Principal Components Analysis (PCA) to describe the PSF shape. Using an analytical approach and an empirical one based on shapelets (\citealt{massey05}), \cite{sph08} have investigated the impact of imperfect shape description of the PSF on dark energy constraints, and set requirements on the number of stars needed to calibrate the PSF. \cite{sph09} explored how the sparsity and the complexity of a PSF model affect the level of systematics in weak lensing surveys. Recognizing that any modeling method --relying on the interpolation of the incomplete information contained by individual stars, each one being an imperfect realization of the PSF-- can be reliable only on scales larger than the typical distance between stars, \cite{amara10} investigated the hardware-software balance: although the PSF can be corrected for on large scales through thorough modeling, it is hardware driven on small scale, requiring that the telescope's characteristics are set such that the PSF is known, stable and well controlled at small scales. \cite{rowe10} has discussed how to diagnose a given PSF model. In particular, he showed how measuring the correlation function of the residuals' ellipticity is an ultimate test to the PSF interpolation.

A first step in PSF modeling is to characterize it where it is sampled (i.e. on stars). Several quantities can be used to describe the PSF properties, such as its moments (e.g. \citealt{ksb,rrg}), or its full shape through shapelets coefficients (\citealt{berge08}) or Principal Components (\citealt{jee07}).
\cite{jee07} have investigated how wavelets, shapelets, and Principal Components perform to measure the stars' shapes. They found that among those three techniques, Principal Components provide the best description of the stars' shapes. We thus follow their method to extract the shape information from stars.

The next step is to interpolate the stars' shape information to the positions of interest, usually galaxies. The goal of this paper is to assess which interpolation scheme performs best to model a realistic PSF field, such as what can be dealt with in weak lensing analyses. To this end, we develop simulations of PSF fields, where stars are built from a set of basis functions defined as the Principal Components of real stars in a Subaru image. We then run our PSF modeling pipeline, including a PCA and an interpolation of the stars' principal components coefficients, with different interpolation schemes.

We shall review the different interpolation schemes that we use in Sect. \ref{sect_interps}. Section \ref{sect_sims} presents the simulations that we create and use to compare the behavior of those interpolation schemes. Our results are shown in section \ref{sect_results}. We discuss them, including the dependence of the interpolation schemes on the stellar density, their sensitivity to outliers, and their impact on weak lensing systematics, in Sect. \ref{sect_discussion}. We conclude in Sect. \ref{conclusion}.

\section{Interpolation schemes} \label{sect_interps}

This section presents the basics of the interpolation schemes compared in this paper. More details can be found in the Numerical Recipes (\citealt{nr}). \cite{platen11} proceed with similar comparisons, in the context of nonlinear density field reconstruction; they give additional useful mathematical background to some techniques used here.

\subsection{Polynomial interpolation}

It is common practice in weak lensing to assume that the PSF varies smoothly across the field, and to interpolate it from stars with a bivariate polynomial (e.g. \citealt{vwaerbeke05,miyazaki07,berge08,fu08}). 
Different variations have been investigated and used on real data: rational fractions (\citealt{vwaerbeke05}), decomposition of the PSF field into subfields corresponding to each chip of the camera to help capture each chip's intrinsic behavior, or superposition of bivariate polynomials (Barney Rowe, private communication).

At the core of polynomial interpolation lies the assumption that the PSF spatial variation can be described by a smooth analytic function (for instance, a polynomial). 
However, there is no strong physical reason why the PSF field should vary as a polynomial (e.g., the intrinsic camera's PSF can be discontinuous from chip to chip). Hence, a simple bivariate polynomial is frequently no more than a rather good approximation of the PSF field, which may not capture details well enough for precision-cosmology.
Moreover, one usually does not need to know such an analytic form, but only needs to interpolate the PSF characteristics at positions of interest (e.g. galaxies). It is therefore completely legitimate to consider the PSF field as a spatial random field, and to turn to less- or non-analytic forms of interpolation, such as those presented below.

\subsection{Radial Basis Function interpolation}

The Radial Basis Function (RBF) interpolation assumes that the spatial variations of the field $y(\mathbf{x})$ can be represented by the superposition of local functions $\phi(r)$, which depend only on the distance to data points $j$, $r=|\mathbf{x}-\mathbf{x_j}|$:
\begin{equation}
y(\mathbf{x})=\sum_{i=0}^{N-1} w_i \phi(|\mathbf{x}-\mathbf{x_i}|)
\end{equation}
where $N$ is the number of data points and $w_i$ are unknown weights, estimated from the value of the fields on data points.

\subsection{Shepard interpolation}
The Shepard interpolation is a special case of a RBF interpolation where $\phi(r) \rightarrow \infty$ as $r \rightarrow 0$. It can be shown that in this case:
\begin{equation}
y(\mathbf{x})=\frac{\sum_{i=0}^{N-1} y(\mathbf{x_i}) \phi(|\mathbf{x}-\mathbf{x_i}|)}{\sum_{i=0}^{N-1} \phi(|\mathbf{x}-\mathbf{x_i}|)}.
\end{equation}
In the following, we test a Shepard interpolation with $\phi(r)=r^{-p}$, with $p>0$.

\subsection{Kriging interpolation}

Kriging is a Gaussian process regression technique which has been shown to provide the best linear unbiased estimator of a statistical field (\citealt{cressie88,cressie93}). 

We assume that we have $N$ data points $\mathbf{x_i}$, where the field $y_i=y(\mathbf{x_i})$ is known, and that we want to estimate the field at a given point $\mathbf{x_G}$. 
Kriging looks for the value of the field at this position as a weighted linear combination of the nearby values at known positions:
\begin{equation}
\hat{y}_G=\sum_{i=0}^{N-1}\lambda_i y_i.
\end{equation}

The weights $\lambda_i$ are estimated such that they minimize the error with respect to the data according to the mean square variation. Therefore, Kriging relies on the estimation of the variogram of the data to interpolate, the variogram being the mean square variation of the values of the field $y(\mathbf{x})$ as a function of the offset distance $\mathbf{r}$:
\begin{equation}
v(\mathbf{r}) \sim \frac{1}{2}\left< [y(\mathbf{x}+\mathbf{r})-y(\mathbf{x})]^2 \right>
\end{equation}
where the average is over all $\mathbf{x}$ and $\mathbf{r}$. In the following, under the assumption of isotropy, we assume that the variogram only depends on the distance $r=|\mathbf{r}|$.

We note $v_{ij}=v(|\mathbf{x_i}-\mathbf{x_j}|)$ and $v_{Gj}=v(|\mathbf{x_G}-\mathbf{x_j}|)$. If we define the vectors
\begin{equation}
\mathbf{Y}=(y_0,y_1,\dots,y_{N-1},0) \,\, \mbox{and}
\end{equation}
\begin{equation}
\mathbf{V_G}=(v_{G,0},v_{G,1},\dots,v_{G,N-1},1)
\end{equation}
as well as the matrix
\begin{equation}
\mathbf{V}=\left(
\begin{array}{ccccc}
v_{00} & v_{01} & \dots & v_{0,N-1} & 1 \\
v_{10} & v_{11} & \dots & v_{1,N-1} & 1 \\
& & \dots \\
v_{N-1,0} & v_{N-1,1} & \dots & v_{N-1,N-1} & 1 \\
1 & 1 & \dots & 1 &0
\end{array}
\right)
\end{equation}
then the Kriging interpolation estimate $\hat{y}_G \approx y(\mathbf{x_G})$ is
\begin{equation}
\hat{y}_G=\mathbf{V_G}\mathbf{V}^{-1}\mathbf{Y}.
\end{equation}

The extra row and column in $\mathbf{V}$, as well as the extra elements in $\mathbf{Y}$ and $\mathbf{V_G}$ allow the estimator to be unbiased.

We find that an exponential variogram, defined as $v(r)=\sigma^2 \exp(-r/a)$, where $\sigma^2$ is the data points' variance and $a$ is a free parameter, gives very competitive results. The free parameter $a$ is the range, which sets how far from the position of interests data points should be used; at distances larger than $a$, the variogram becomes constant. Another free parameter can be added, called the nugget, which describes the degree of correlation of the field at very small scales, $v_0=\lim_{r\rightarrow 0}v(\mathbf r)$; in this work, we find that setting $v_0=0$ (i.e., assuming the field does not decorrelate at very small scales) gives the best results. 

\subsection{Delaunay triangulation}
Delaunay triangulation is a particular type of triangulation. Drawing triangles whose vertices are the data points, the Delaunay triangulation is that which maximizes the triangles' angles and minimizes their sides' length. The function to be interpolated is known at the triangles' vertices, then interpolated in their interior, where it must be estimated. In this paper, we use a linear interpolation between the triangles' vertices.

\section{Simulations} \label{sect_sims}
In order to test the different interpolation schemes presented in section \ref{sect_interps}, we developed versatile simulations. They create mock PSF fields based on real images, whose spatial variation, star density and star signal-to-noise ratio (S/N) can be set.

Our goal is to create mock PSF fields that are as realistic as possible, and can be tailored to images from current surveys. We thus base our simulations on real images, that we call `reference images'. 
To create a simulated image, we first extract the stars from the reference image using SExtractor (\citealt{sextractor}). We define a set of basis functions, with which a mock star can be created, by proceeding to a Principal Components Analysis of the stars from the real image; the associated Principal Components (PC) eigenvectors define the set of basis functions. During this process, we keep 90\% of the variance information of the stars, so that the shape information is well extracted, while the background is discarded.

We then make simulated stars by adding linearly the weighted PC eigenvectors. To obtain various shapes of stars, the weights are chosen randomly, with the only constraint that the variance of any given weight is similar to that of the coefficient associated with the corresponding PC eigenvector of the real stars, to ensure that mock stars are realistic.  Finally, we require that the PSF varies across the image. We set its spatial variation by defining the spatial variation of each weight of the PC reconstruction: the random field from which we draw the weights is therefore given a realistic power spectrum, following the technique described by \cite{rowe10}.

For the sampling of the PSF field, i.e. the position of $n_{\rm stars}$ stars across the image, we can either (1) use the position of stars of the real image (in this case, $n_{\rm stars}$ is the number of stars in the real image) or (2) lay any number $n_{\rm stars}$ of stars at random positions. The former possibility is especially useful to diagnose the PSF model that we can expect to obtain when analyzing the real image; for instance, any sub-populated area may suffer from a badly constrained model, which will be revealed when analyzing the mock. The latter is useful for general modeling method development, e.g. to test the behavior of a given method on the stellar density. We use this option in this paper.

Besides stars, we simulate $n_{\rm gals}$ extra PSFs at random positions, that we call `galaxy-PSFs' in the remainder of this paper. They are at the positions at which we want to interpolate the PSF from the $n_{\rm stars}$ above. Simulating those PSFs, instead of only storing their characteristics, allows us to have a realistic image, on which we can run our PSF measurement pipeline in the same way as we do on real data (that is, there is no need to tailor it to the simulations), as well as to directly compare the PSF image and the model, and to use the exact same techniques to measure the PSF's characteristics on the simulated image and on the modeled image.

A background with the same statistics as that of the reference image is then added to the simulated image.
The S/N distribution of simulated stars is set by hand, either following that of the real image, or independent of it, to test the behavior of the interpolation technique on stars' S/N.

\begin{figure}
\centering
\includegraphics[width=8cm,angle=0]{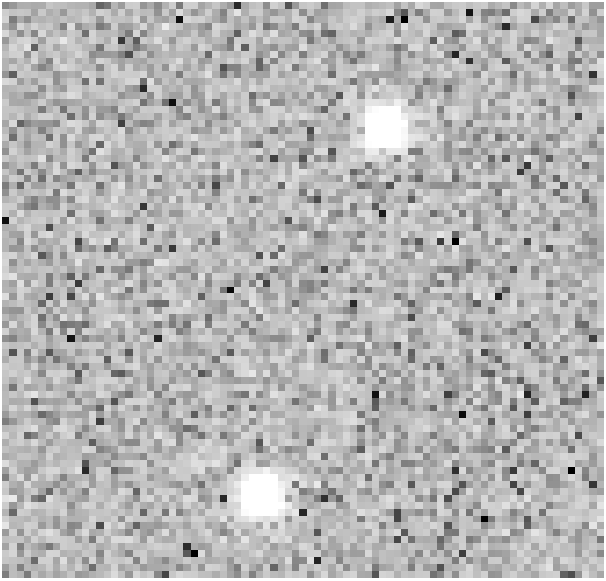} 
\caption{Example of PSFs in one simulation, with a logarithmic color scale. The pixel scale is 0.2 arcsec/pixel.} \label{fig_sim}
\end{figure}

In this paper, we base our analyses on 50 mock Subaru images, with a reference image being a typical Subaru SuprimeCam image. The simulations are 10,032$\times$7,769 pixels$^2$ in area, corresponding to 33$\times$26 arcmin$^2$. The PC decomposition of the stars in the reference image provides us with a set of 13 basis functions, which are used to make simulated stars. We set the spatial variation of the simulated stars' PC weights with a power spectrum of slope 11/4, which is close to what is measured for ground-based PSFs from the Kolmogorov spectrum for atmospheric turbulence (e.g. \citealt{sasiela94}). The resulting PSF's ellipticity is of average 0.05, and of standard deviation 0.02, consistent for all simulations.
Each simulation contains 1000 stars (where the PSF would be sampled for a real image's analysis) randomly distributed, all with the same S/N=100. This stellar density of 1.1 star per arcmin$^2$ is close to what is usually measured in real surveys. `Galaxy-PSFs', where the PSF must be interpolated, are added in two versions of each simulation: one version features randomly positioned `galaxy-PSFs', while `galaxy-PSFs' of the other version are on a regular grid. The latter version allows us to use powerful 2D plots; the former is closer to reality and is used to measure correlation functions. We input 2000 `galaxy-PSFs' in each version of each simulation. In the following, only those `galaxy-PSFs' are used to assess the correctness of any interpolation.
Figure \ref{fig_sim} shows two PSFs in one of our simulations. The left panels of Fig. \ref{fig_ell} show a PSF ellipticity field on the `galaxy-PSFs' grid (top) and on `stars' (bottom).

\subsection{Quantities of interest}

We consider two quantities of interest to characterize the simulated PSF and test the model: Principal Components coefficients and Gaussian-weighted ellipticities 
\begin{equation}
e=e_1+{\rm i} e_2=\frac{Q_{11}-Q_{22}+2{\rm i}Q_{12}}{Q_{11}+Q_{22}},
\end{equation}
where $Q_{i,j}$, ($i,j=1,2$) are the PSF's quadrupoles.

To compare the modeled PC coefficients with the input ones, we force our PCA-PSF measurement pipeline to use the exact same set of PC as those used for the simulation. This downgrades the capacity of our pipeline, since it is not free to find the PC eigenvectors' set  that optimally describes the PSF. However for high enough S/N stars, this has a negligible effect. Moreover, when defining mock stars from the reference image's PC eigenvectors, we are careful to normalize the stars in the same way as we do in our PCA-PSF measurement pipeline, to ensure that PC keep the same meaning during the entire process from simulation to modeling of the PSF to test of the model. Therefore, it makes sense to compare the interpolated PC coefficients with those input in the simulations.

\section{Results} \label{sect_results}

\begin{figure*}
\centering
\includegraphics[height=21cm,angle=180]{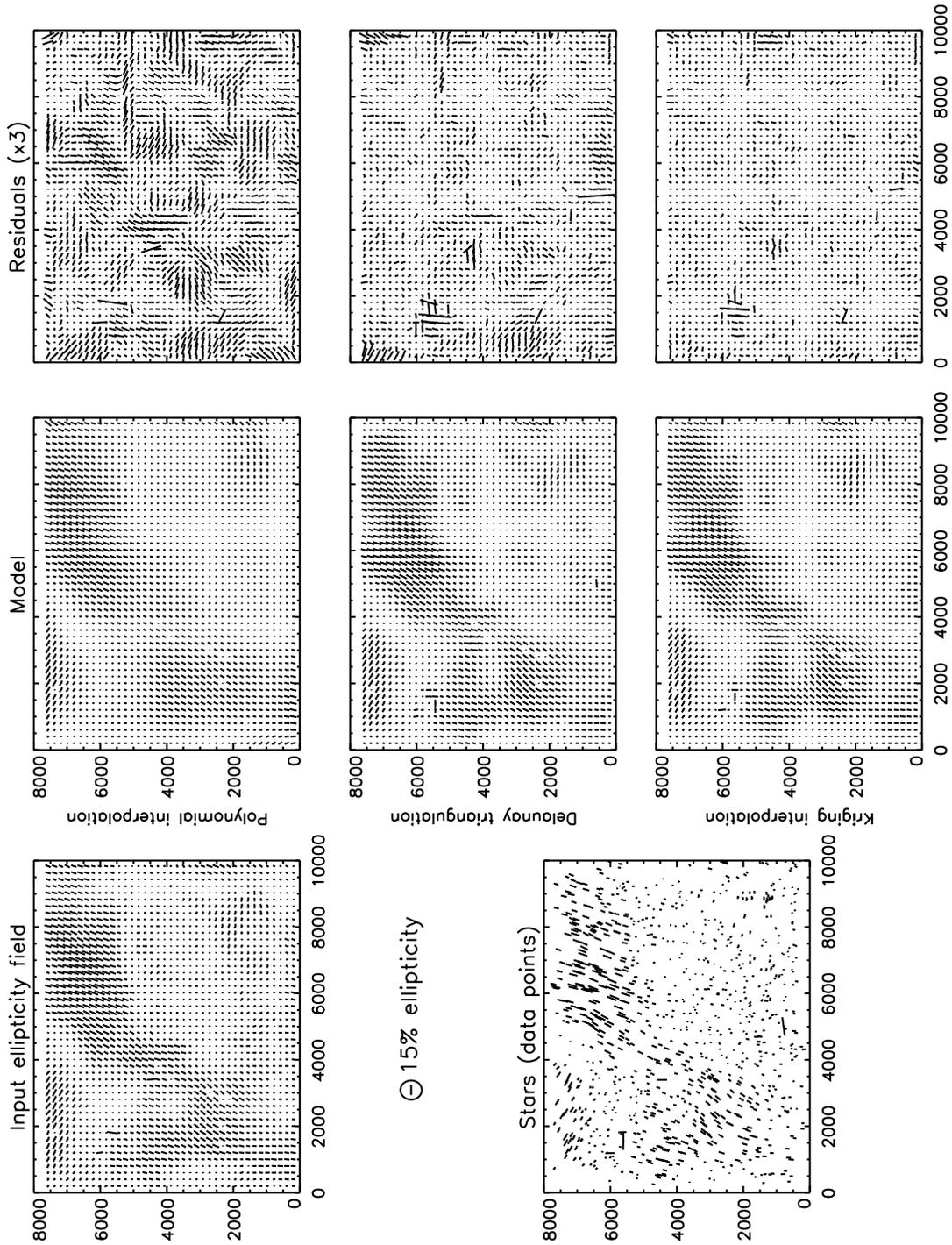}
\caption{Unweighted ellipticity of the simulated PSF field. Top-left: ellipticity of the input PSF on a regular grid. Bottom-left: ellipticity of the input PSF at the position of `stars' used as samples of the PSF for the interpolation of its Principal Components. Middle column: recovered PSF ellipticity fields after interpolation of the PSF PCs, for polynomial interpolation (top), Delaunay triangulation (center) and Kriging interpolation (bottom). Right column: residuals, amplified by a factor 3 for better visibility.} \label{fig_ell}
\end{figure*}

\begin{figure*}
\centering
$ \begin{array}{c}
\includegraphics[width=11cm,angle=90]{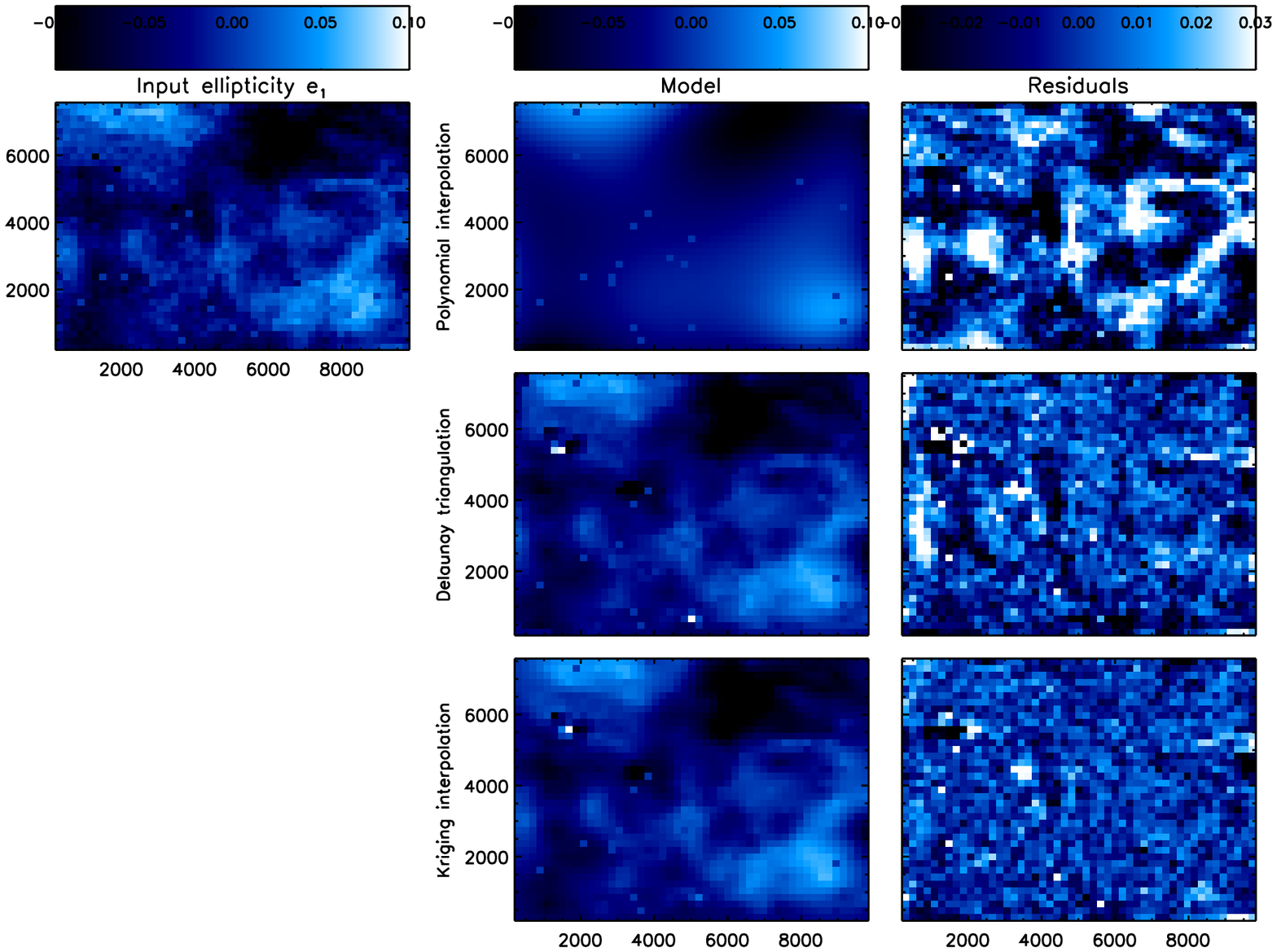} \\
\includegraphics[width=11cm,angle=90]{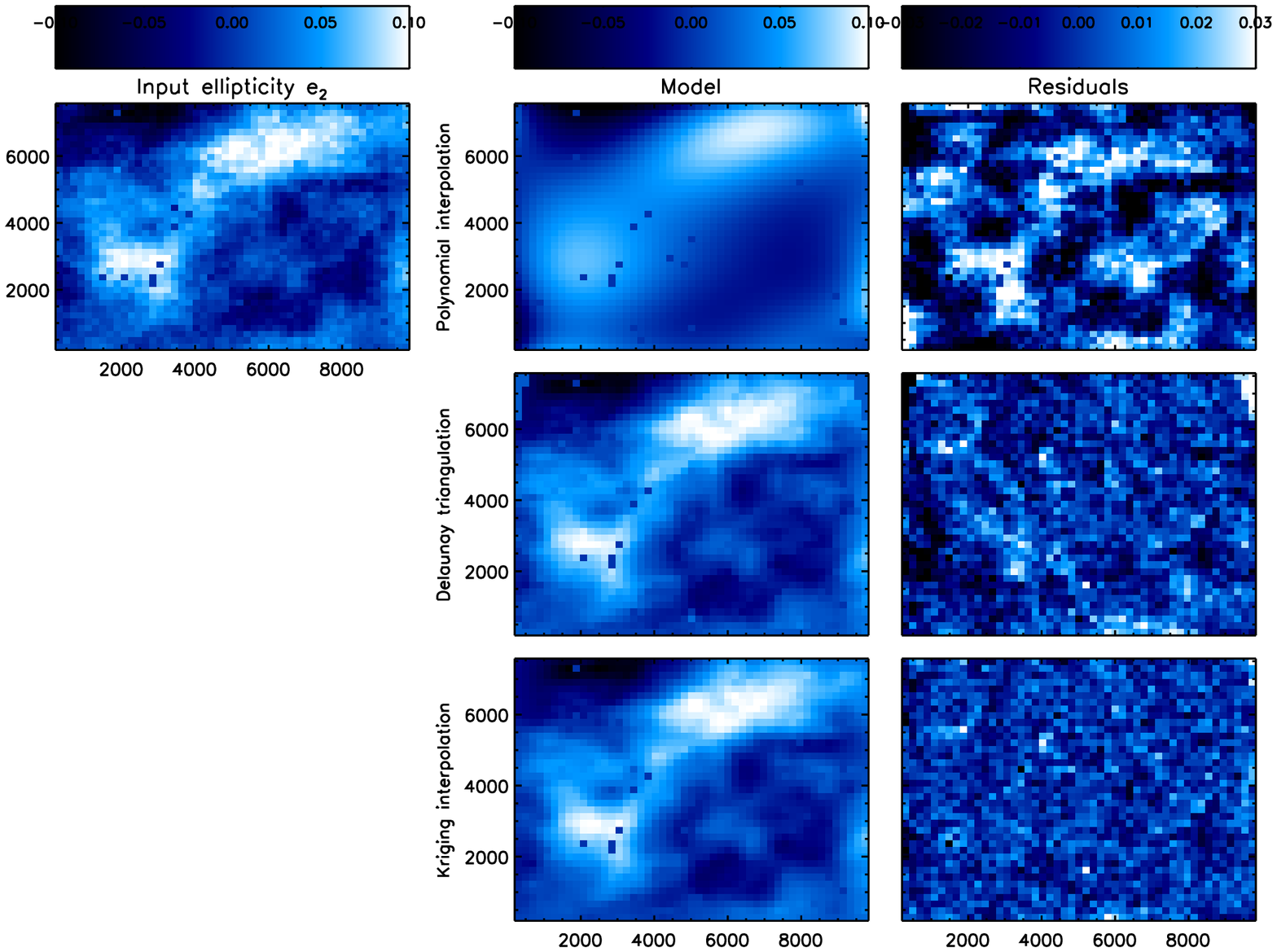}
\end{array} $
\caption{Same as Fig. \ref{fig_ell}, for the two components of the unweighted ellipticity of the PSF. Top: $e_1$. Bottom: $e_2$. Due to gridding aspects, it was not possible to represent the ellipticity components for individual stars. Note that the color scale for the residuals is narrower than that of the input and of the model (-0.03 to 0.03 instead of -0.1 to 0.1).} \label{fig_ell1comp}
\end{figure*}

\begin{figure*} 
\centering
\subfigure{
\label{fig_coeff12}
$ \begin{array}{c}
\includegraphics[width=11cm,angle=90]{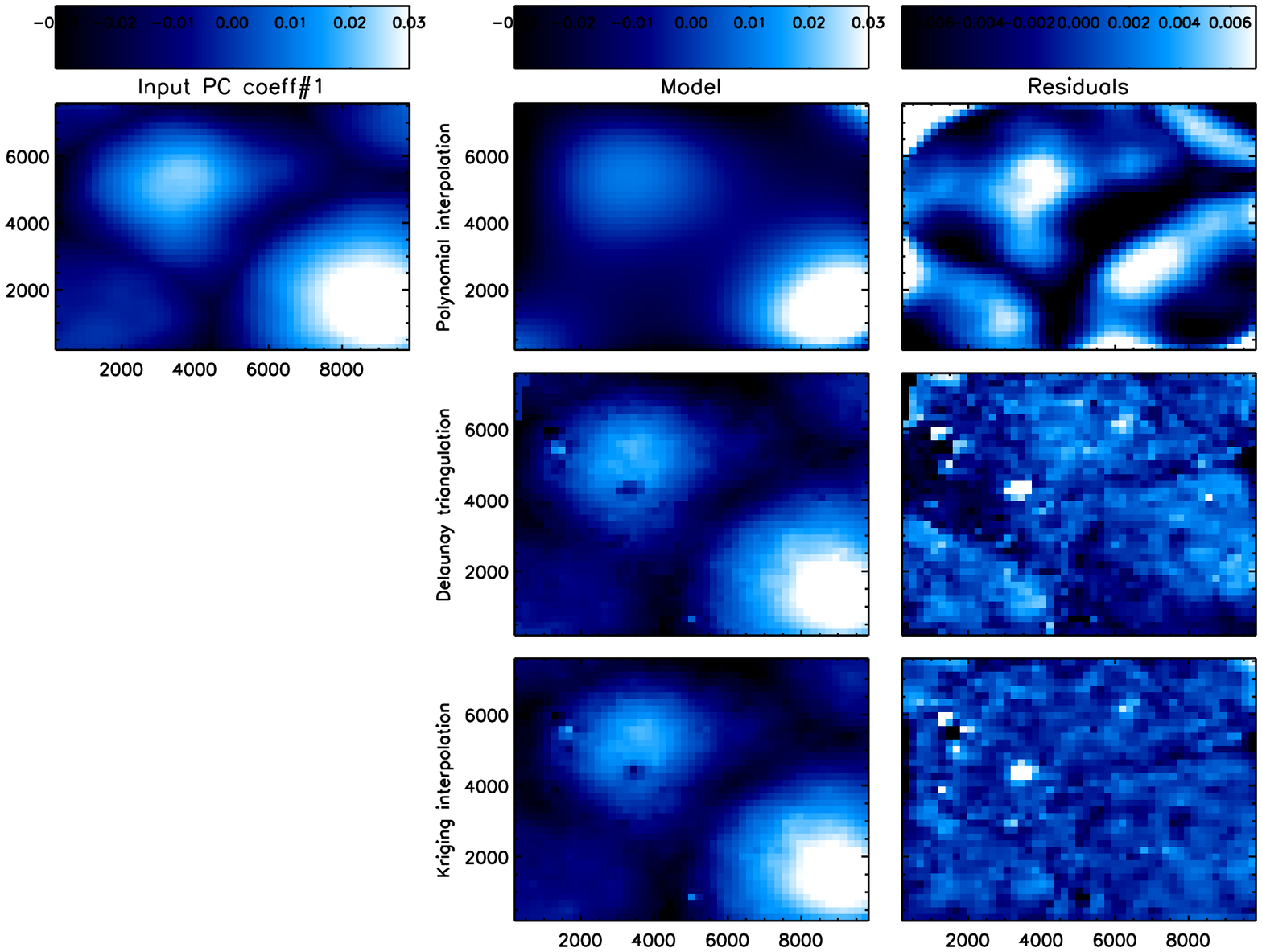} \\
\includegraphics[width=11cm,angle=90]{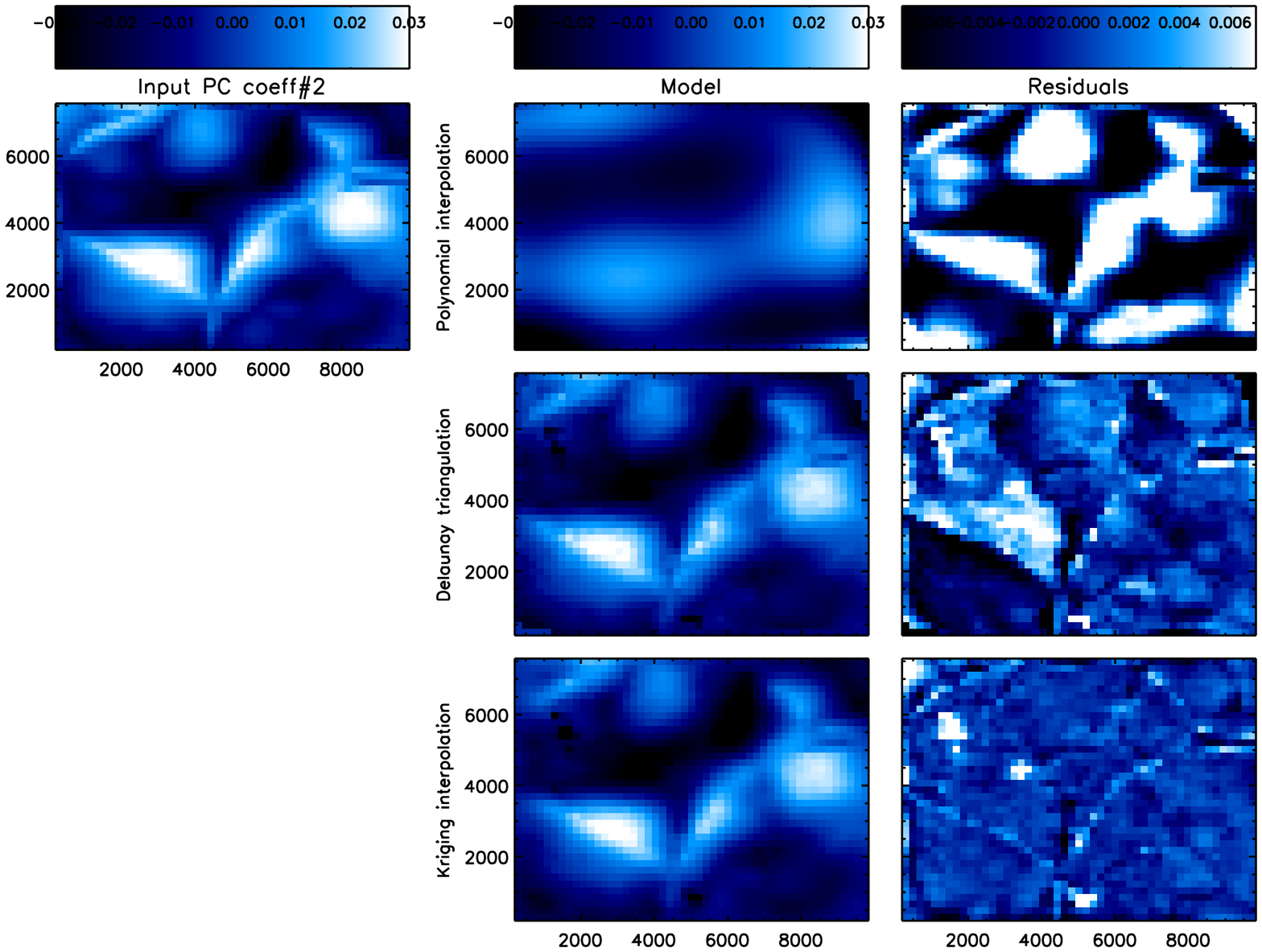}
\end{array} $
}
\caption{\small Same as Fig. \ref{fig_ell}, for the first two Principal Components coefficients of the PSF. Note that the color scale for the residuals is narrower than that of the input and of the model (-0.006 to 0.006 instead of -0.03 to 0.03).}
\label{fig_coeff0}
\end{figure*}

We run our PSF modeling pipeline on the simulations presented in the previous section. As mentioned above, we force our PCA analysis to use the same PC eigenvectors as those used to create the simulated stars. Doing so, our PCA decomposition of an infinite S/N star would output the exact input star.  
Hence, two sources of error are present when we compare the interpolated PSFs with those input in the simulations: the error coming from the shape measurement of finite S/N stars, and that of the interpolation. In this paper, we are concerned with estimating the latter. Having high enough S/N stars, we checked that the errors from the shape measurement are small enough that they can be neglected in our first tests (Fig. \ref{fig_ell}-\ref{fig_coeffscomp}). The last test, using correlation functions (sect. \ref{ssect_corfct}) is sensitive to the errors from the shape measurement; we will use them to discuss the errors coming from the interpolation.

Our results are presented below.

\paragraph*{Polynomial interpolation} We interpolate the PSF's PC coefficients with: (1) a bivariate polynomial on the entire field, the degree of which is focused with a $\chi^2$ minimization, but is constrained to remain small enough to avoid fast divergences; and (2) 2-dimensional Chebyshev polynomials interpolation.

We find Chebyshev polynomials, although they are bound and should be expected to give better results, not to perform better than regular bivariate polynomials. Therefore, in the remainder of this paper, we will only report results from regular bivariate polynomials. We find the best fitting polynomial to be of 5th order.

\paragraph*{RBF interpolation} We try Gaussian-RBF, multi\-qua\-dra\-tic-RBF, inverse multiquadratic-RBF, and thin-plate-RBF. Although Gaussian-RBF behave better than the others, we find all those kinds of RBF to be too unstable, and need too fine a tuning to give reliable models. Therefore, we do not spend more time on RBF interpolations here. However, we find the Shepard interpolation to perform as well as a polynomial interpolation, with almost no fine tuning needed, for $p \geqslant 3$.

\paragraph*{Kriging interpolation} We try exponential and spherical variograms, with varying range and nugget. We find the exponential variogram to be more stable and to perform better than the spherical variogram. Therefore, in the remainder of this paper, we will only consider an exponential variogram when using Kriging interpolation.

We get better results when setting the nugget to 0. When normalizing the images' coordinates so that $0<x<1$ and $0<y<1$, we find that the interpolation gives better results when setting the range of the variogram between $0.25$ and $0.75$. The range acts on the stability of the model at small scales by taking more or less stars into account to estimate the variogram.

\bigskip
In the following, we compare the results of the interpolation with a polynomial, a Delaunay triangulation, and a Kriging technique.


\addtocounter{figure}{-1}
\begin{figure*}
\addtocounter{subfigure}{1}
\centering
\subfigure{
\label{fig_coeff34}
$ \begin{array}{c}
\includegraphics[width=11cm,angle=90]{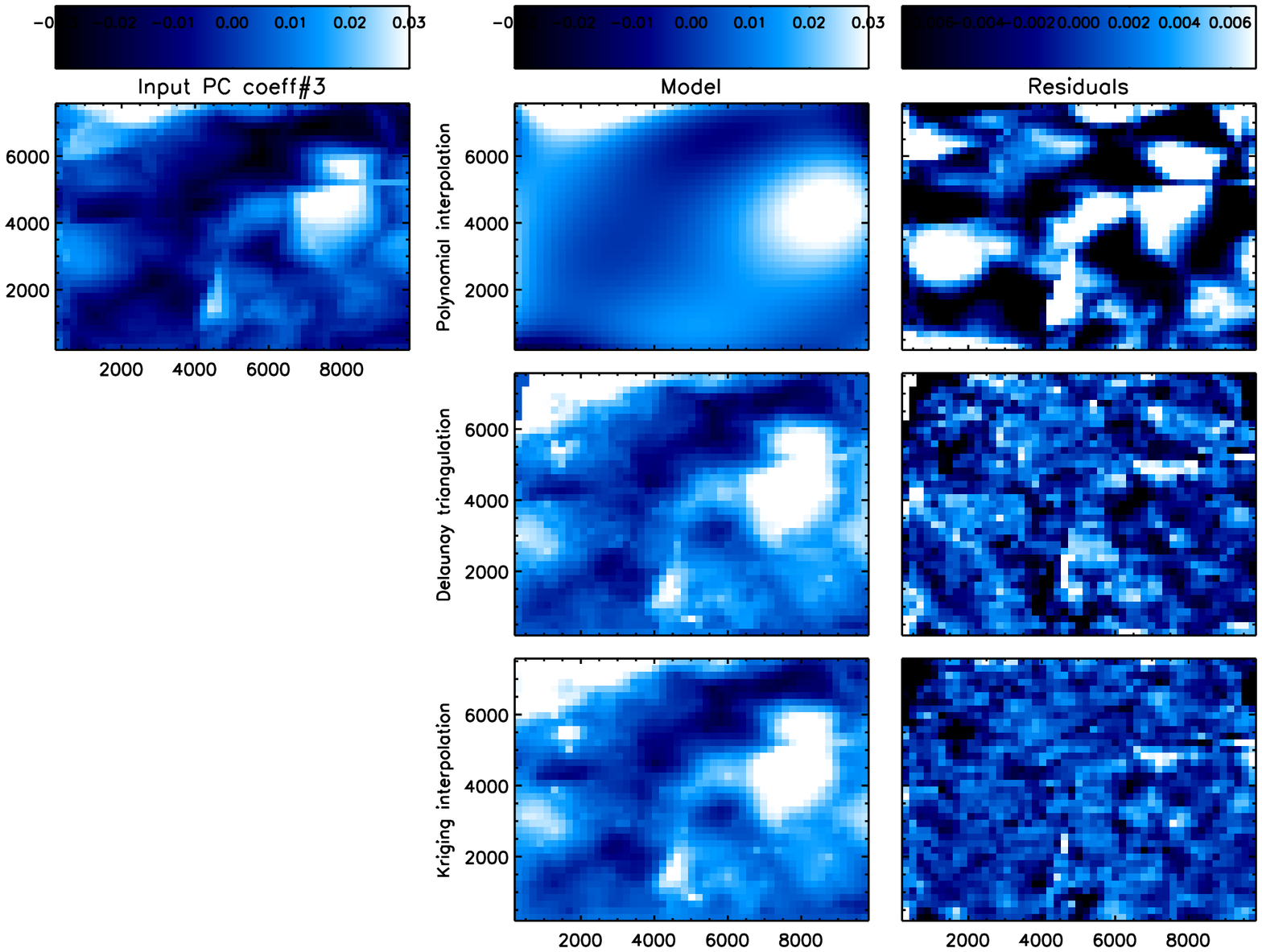} \\
\includegraphics[width=11cm,angle=90]{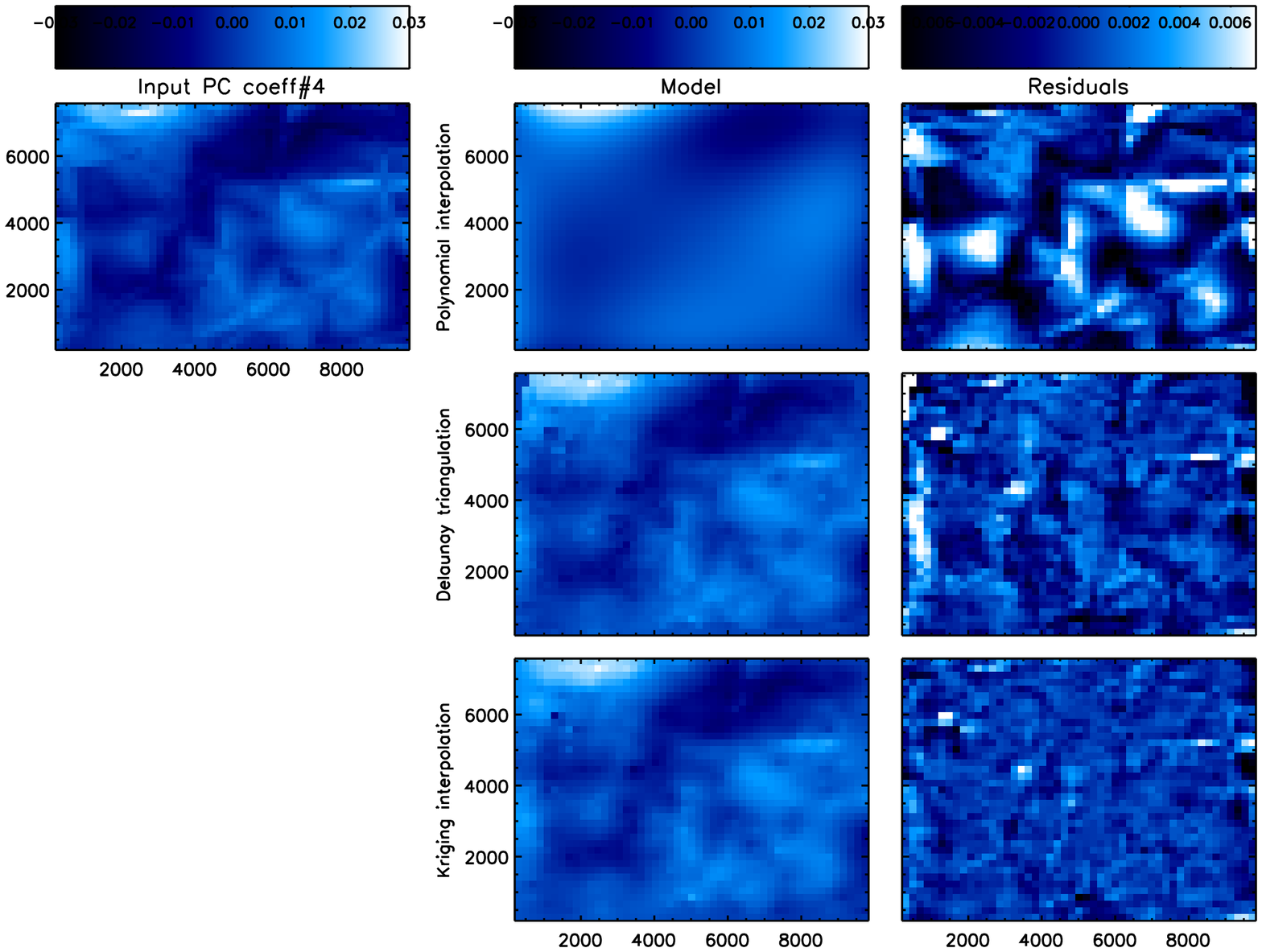}
\end{array} $
}
\caption{\small {\bf (Ct'd)} Third and fourth Principal Components coefficients of the PSF.}
\end{figure*}

\subsection{Recovery of PSF's field information}

We start by comparing the Gaussian-weighted ellipticity of the interpolated PSFs with that of the simulated PSF, for a given simulation. Although the limited space prevents us from showing the results for several simulations, we verified that our results are consistent for all our simulations. The top-left panel of Fig. \ref{fig_ell}, as seen in landscape, shows the simulation's ellipticity field, sampled at the position of `galaxy-PSFs'. The bottom-left panel show the PSF's ellipticity of the stars used for the interpolation.
The middle and right columns of the figure show the models and residuals we obtained with a 5th-order-polynomial (top), Delaunay (middle) and Kriging (bottom) interpolations, on the grid formed by `galaxy-PSFs'.

Figure \ref{fig_ell1comp} shows the same information as Fig. \ref{fig_ell}, decomposed into the two components of the ellipticity, $e_1$ (top) and $e_2$ (bottom). For both components, the top-left panel shows the input, and the right two columns
 show the ellipticity of the interpolated PSF, as well as the residuals, for the three interpolation techniques we compare here.

Figure \ref{fig_coeff0}, with the same layout as Fig. \ref{fig_ell1comp}, goes one step further in decomposing the information. It shows the interpolation of the first four Principal Components (the most significant) of the PSF themselves. Hence, while the previous figures showed the combined `physical' information of the ellipticity, this figure is at the core of our interpolation pipeline, since it shows how each individual interpolation performs. 
Similar figures, made from regularly gridded `galaxy-PSFs' simulations, cannot be done for real data, where one knows the PSF information only on stars. On real data, one can only rely on testing the interpolation of the ellipticity on stars (i.e. one should transform Fig. \ref{fig_ell} such that information appears at the position of stars only).

A bivariate polynomial interpolation fails to capture the quickly varying features of the PSF pattern: significant residuals are seen by eye around the most significant gradients of the PC and ellipticity fields, showing how this interpolation mostly smoothes the information. Higher and higher order polynomials should manage to capture increasingly small variations of the field; however, this requires an uncomfortable tuning of the polynomial, which is hardly compatible with the automatization that is necessary to deal with big surveys.
The Delaunay triangulation, although suffering non-negligible residuals when the field varies quickly (see e.g. the third PC), produces satisfactory interpolations. 
Finally, the Kriging technique recovers the information almost perfectly, with homogeneous, near-zero residuals. Only when the gradients of the fields are very strong does it leave an imprint in the residuals maps.

From these three figures, we can already conclude that the Kriging interpolation performs best, slightly better than a Delaunay triangulation interpolation, and significantly better than a bivariate polynomial interpolation. 
Moreover, the Delaunay triangulation and Kriging interpolations need only minimal tuning, and hence are more portable and adapted to diverse data sets.

\begin{figure}
\centering
$ \begin{array}{c}
\includegraphics[width=8cm,angle=0]{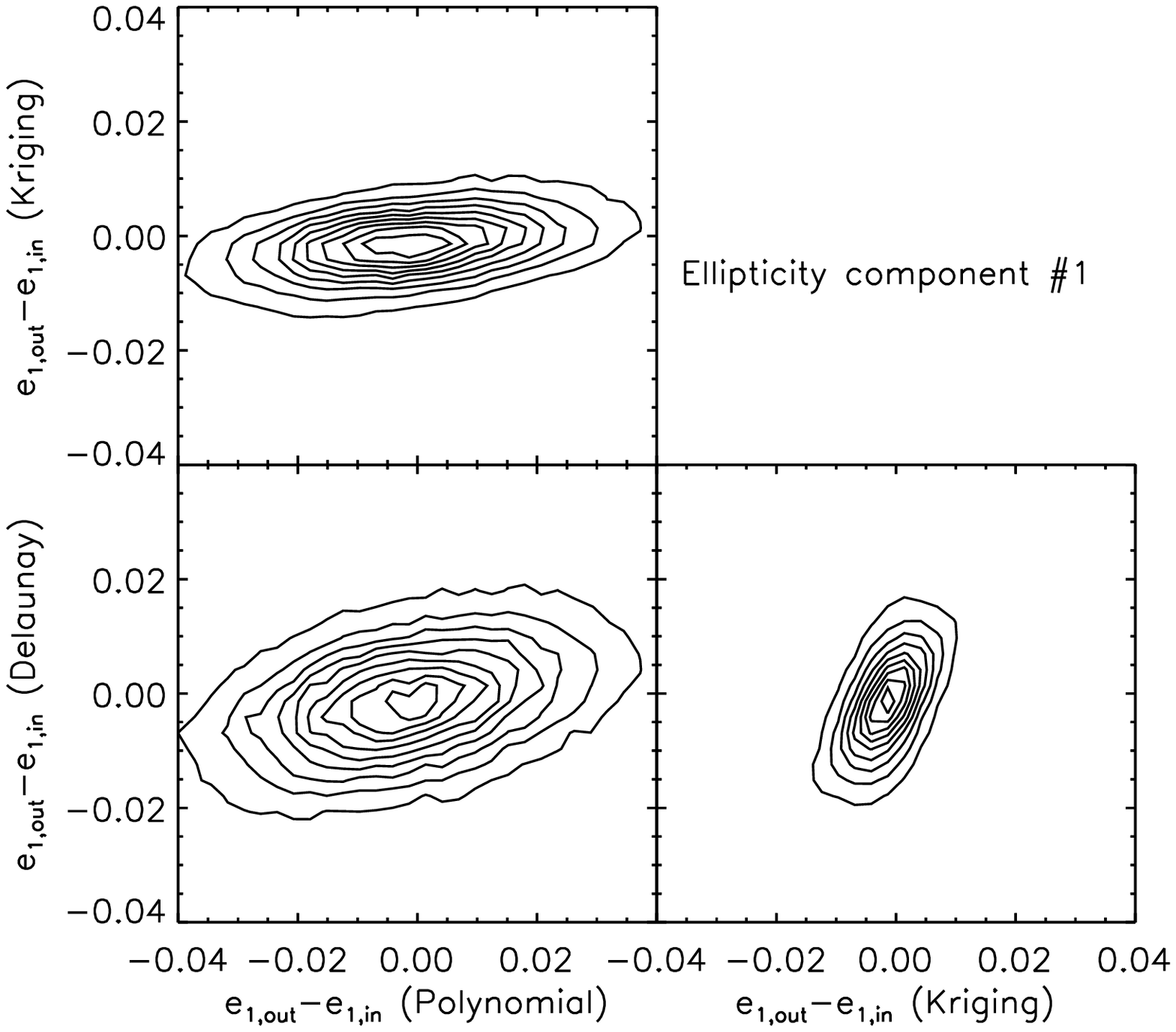} \\
\includegraphics[width=8cm,angle=0]{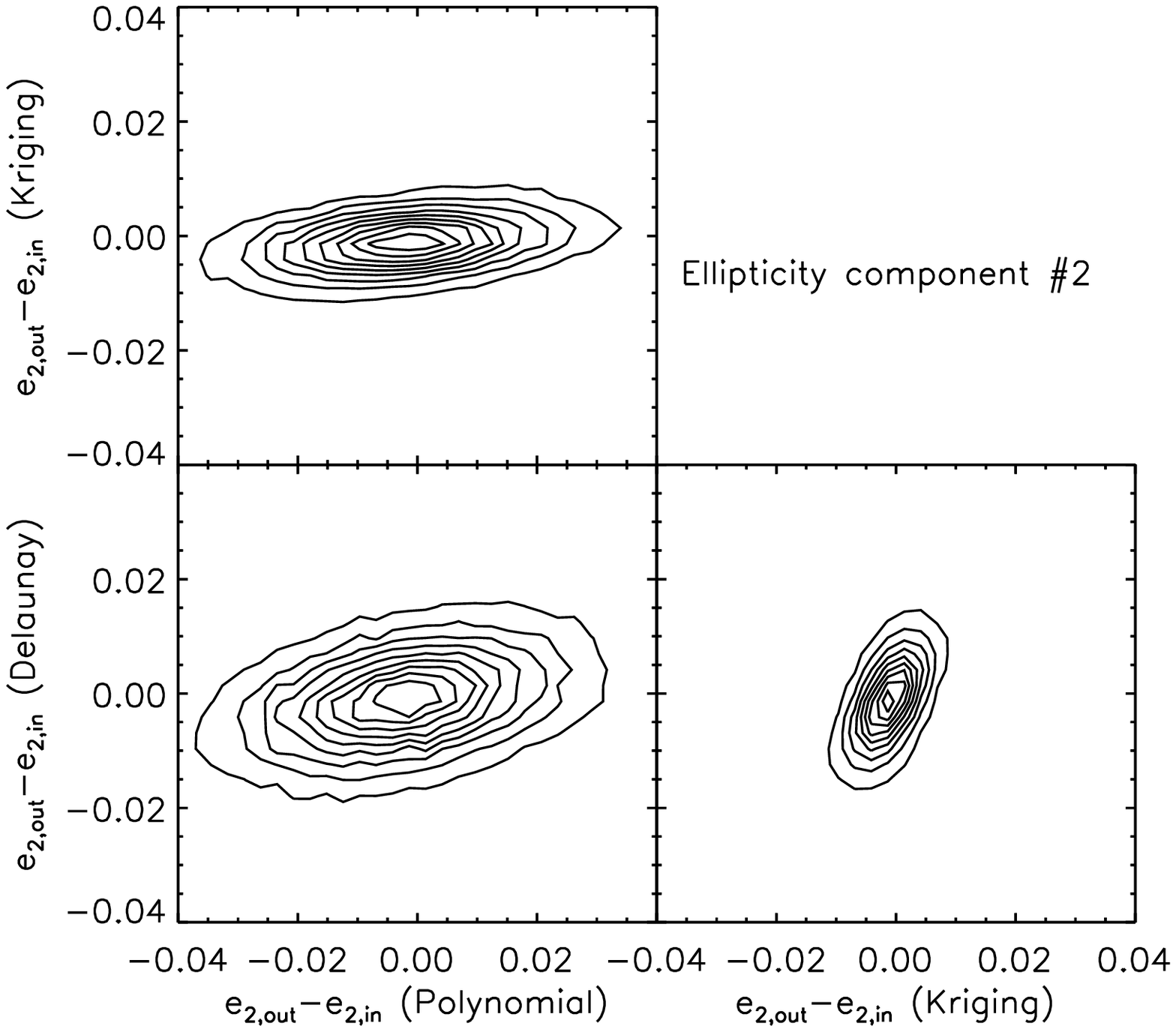}
\end{array} $
\caption{2D distributions of the ellipticity's residuals in the three planes defined by the one-to-one comparisons of the three interpolation schemes. Upper panel: $e_1$. Lower panel: $e_2$. Contours show the distribution of the residuals, normalized such that the maximum is 1. Contours start from 0.1 and increase by steps of 0.1.} \label{fig_ecomp}
\end{figure}
\begin{figure*}
\centering
$ \begin{array}{cc}
\includegraphics[width=8cm,angle=0]{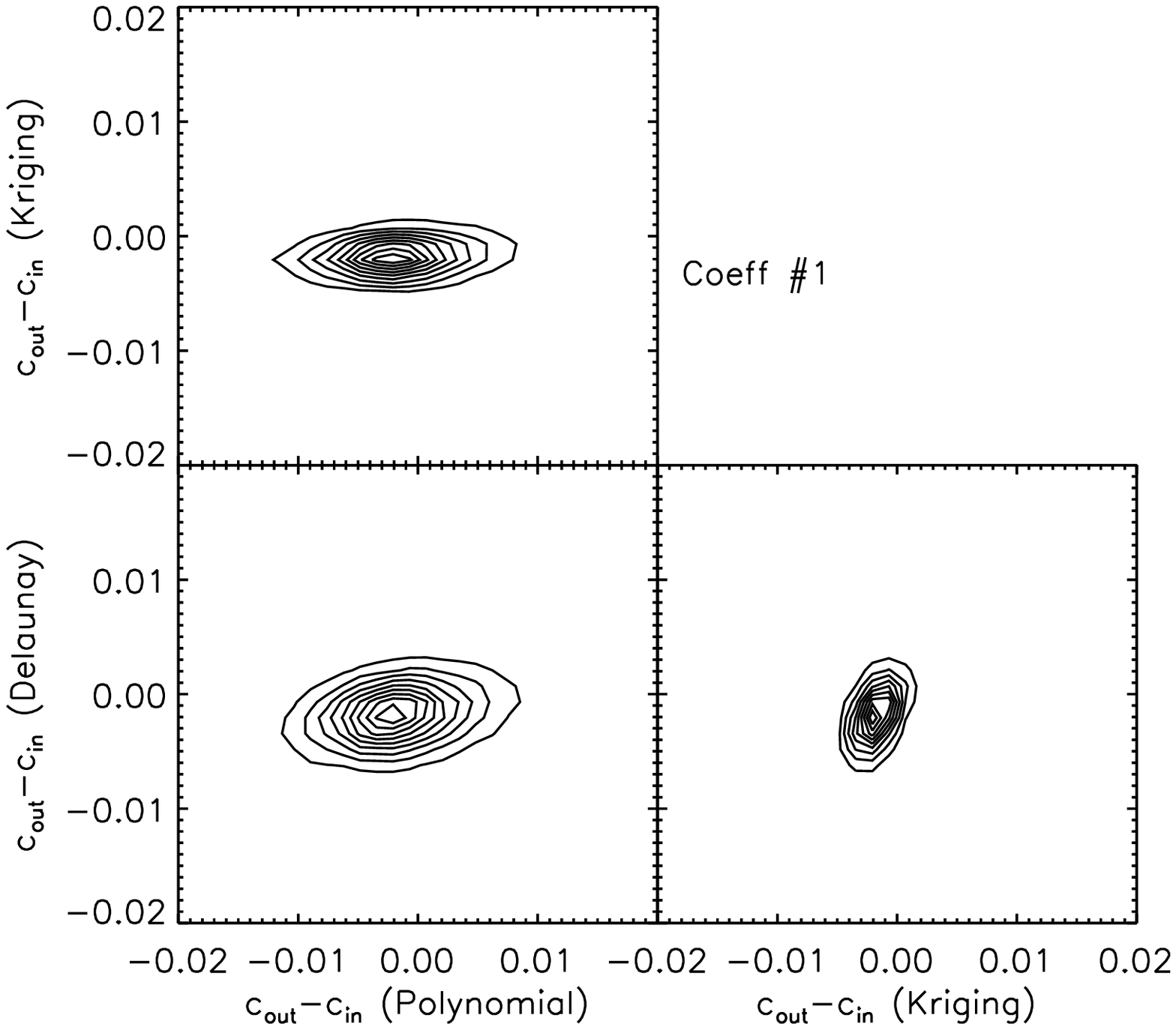}
\includegraphics[width=8cm,angle=0]{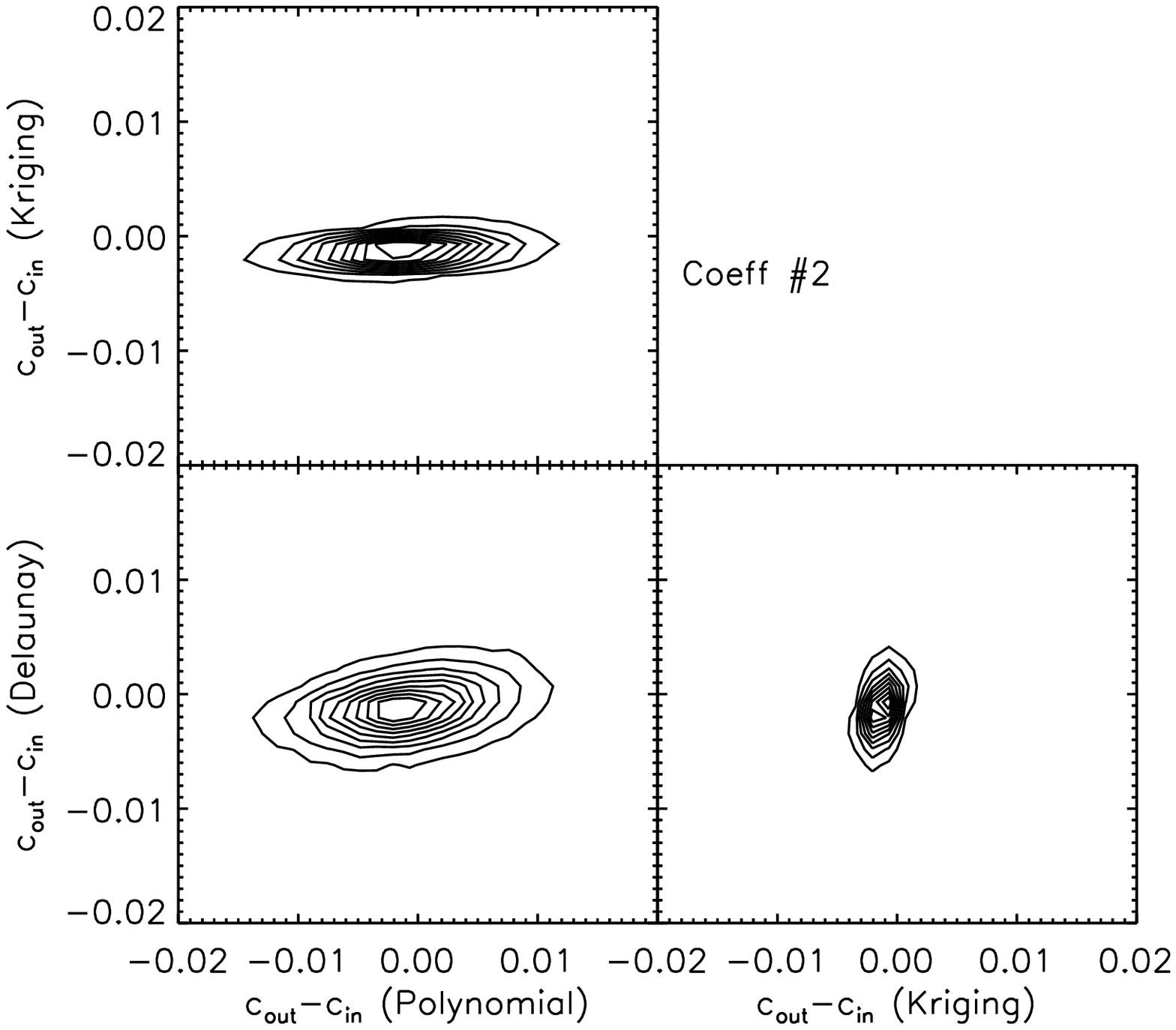} \\
\includegraphics[width=8cm,angle=0]{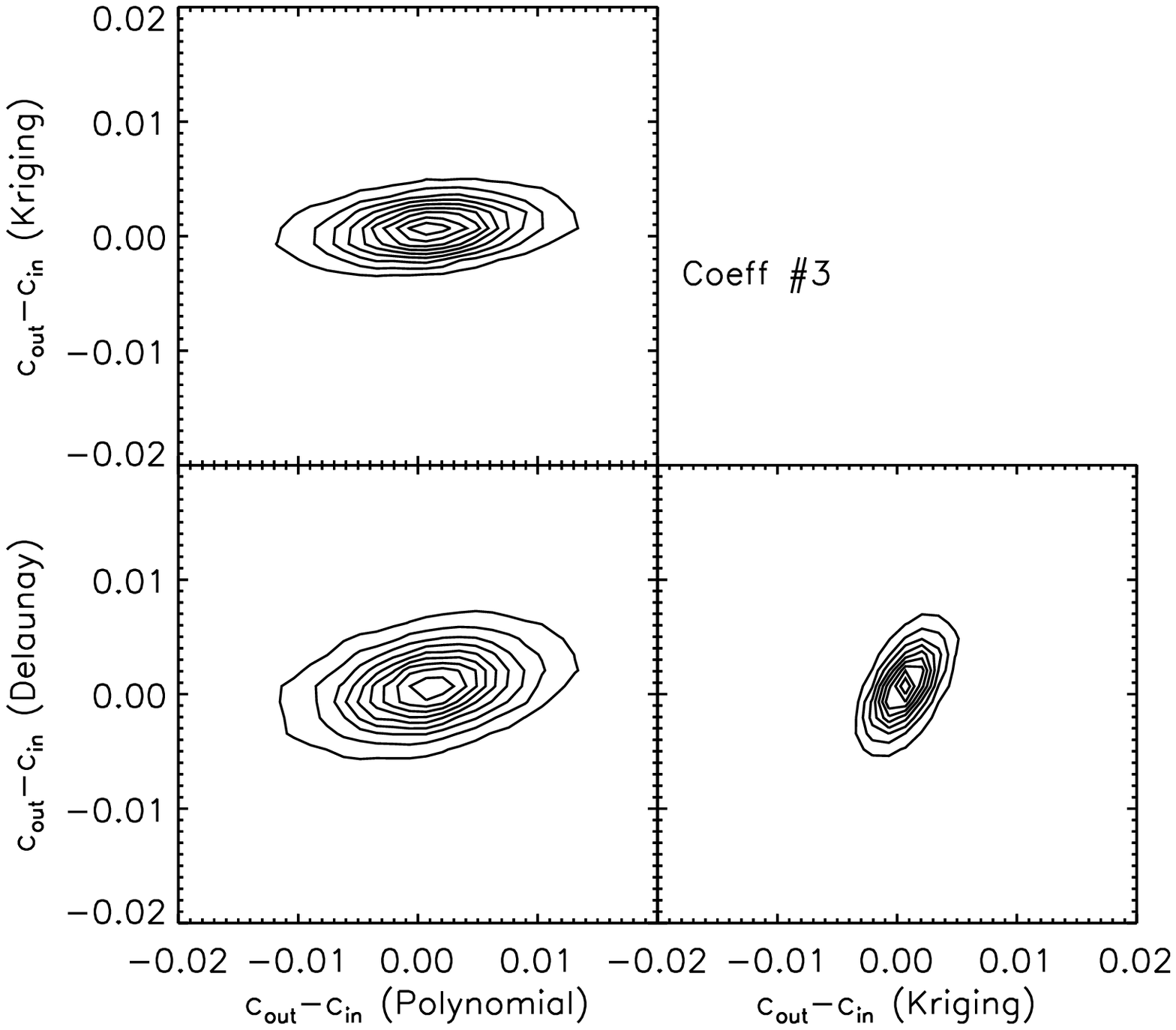}
\includegraphics[width=8cm,angle=0]{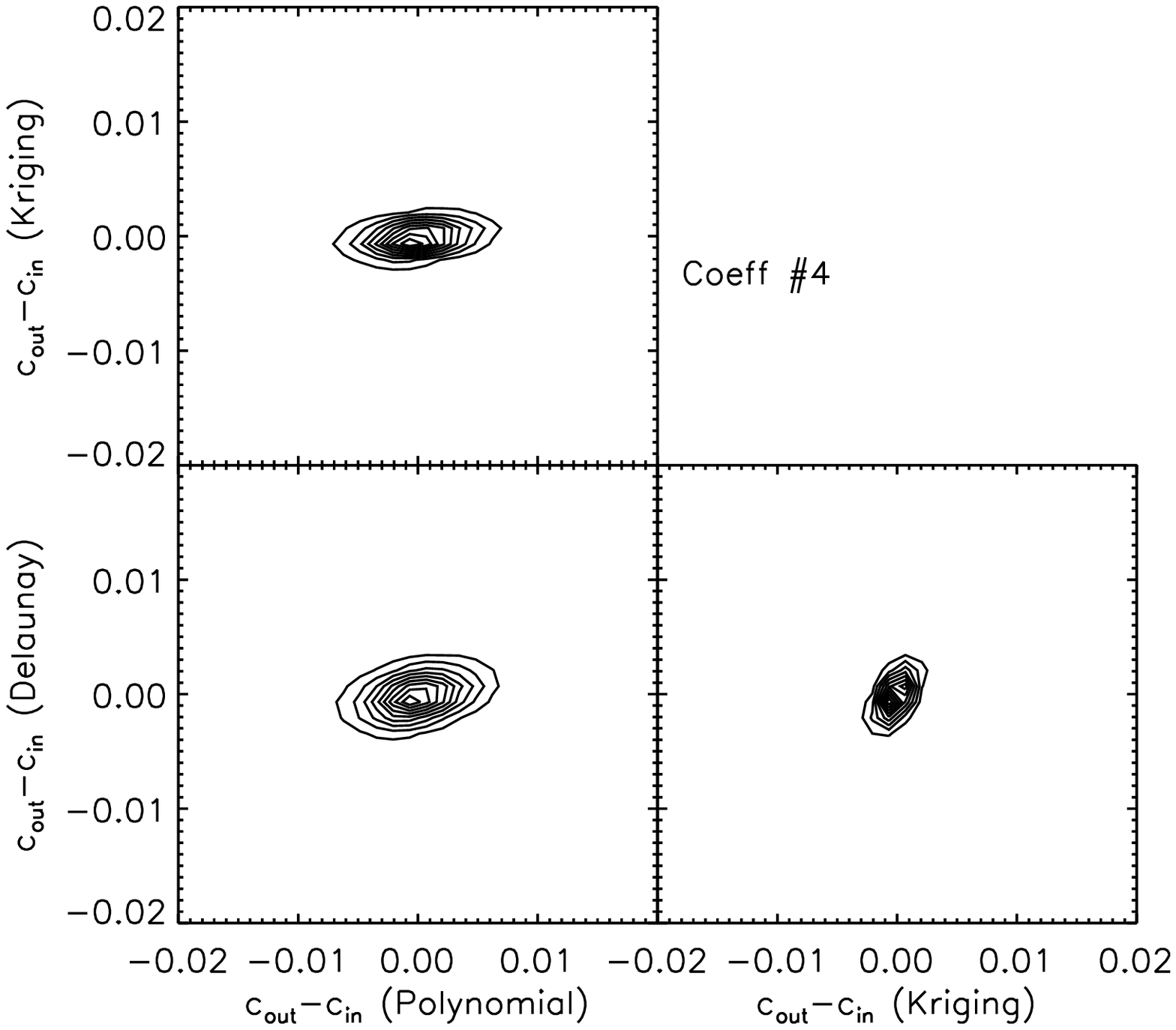}
\end{array} $
\caption{2D distributions of the first four PC coefficients' residuals in the three planes defined by the one-to-one comparisons of the three interpolation schemes. From left to right and top to bottom: first, second, third and fourth coefficient. Contours show the distribution of the residuals, normalized such that the maximum is 1. Contours start from 0.1 and increase by steps of 0.1.} \label{fig_coeffscomp}
\end{figure*}

\begin{figure*}
\centering
\includegraphics[width=6cm,angle=0]{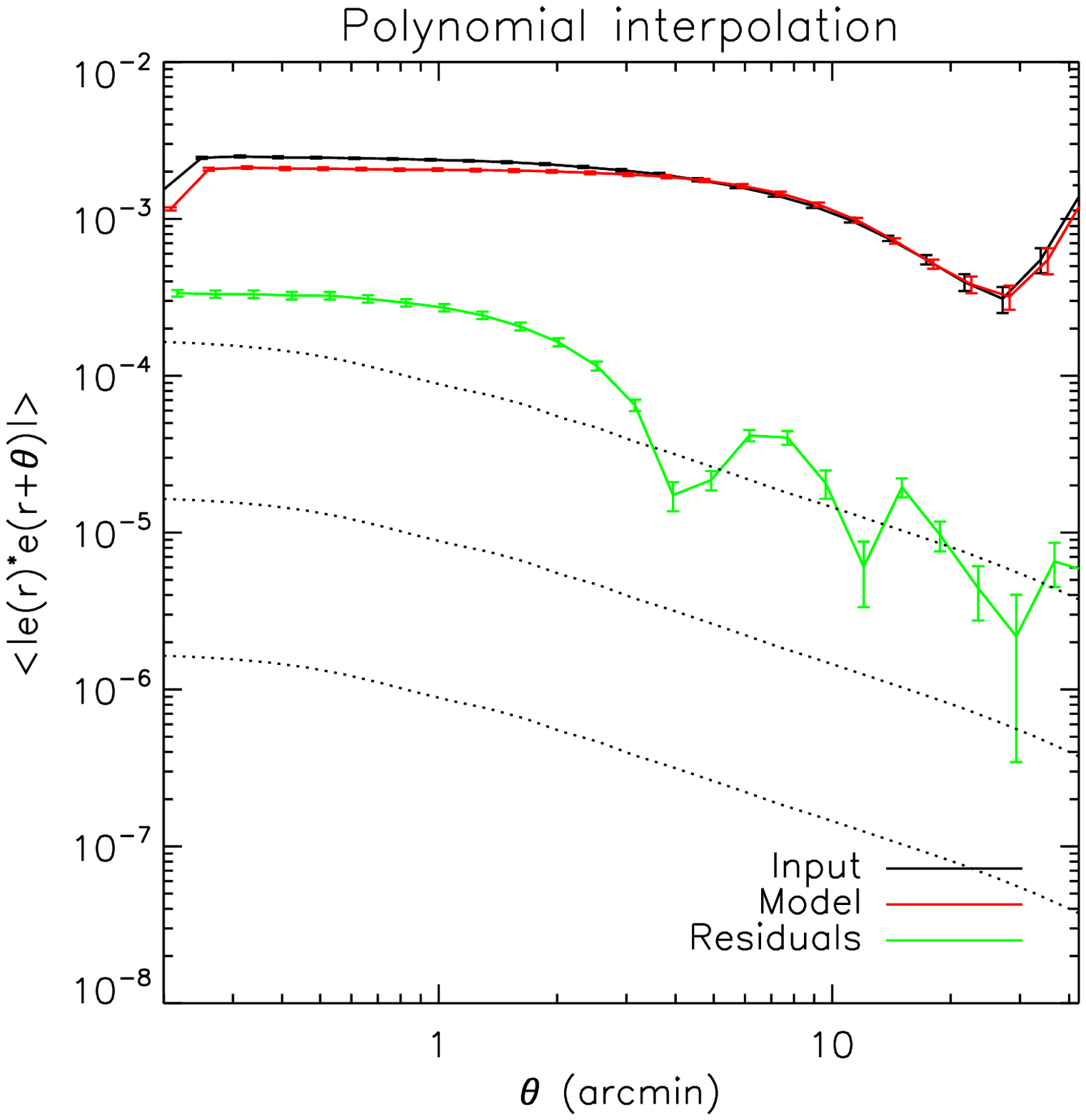}
\includegraphics[width=6cm,angle=0]{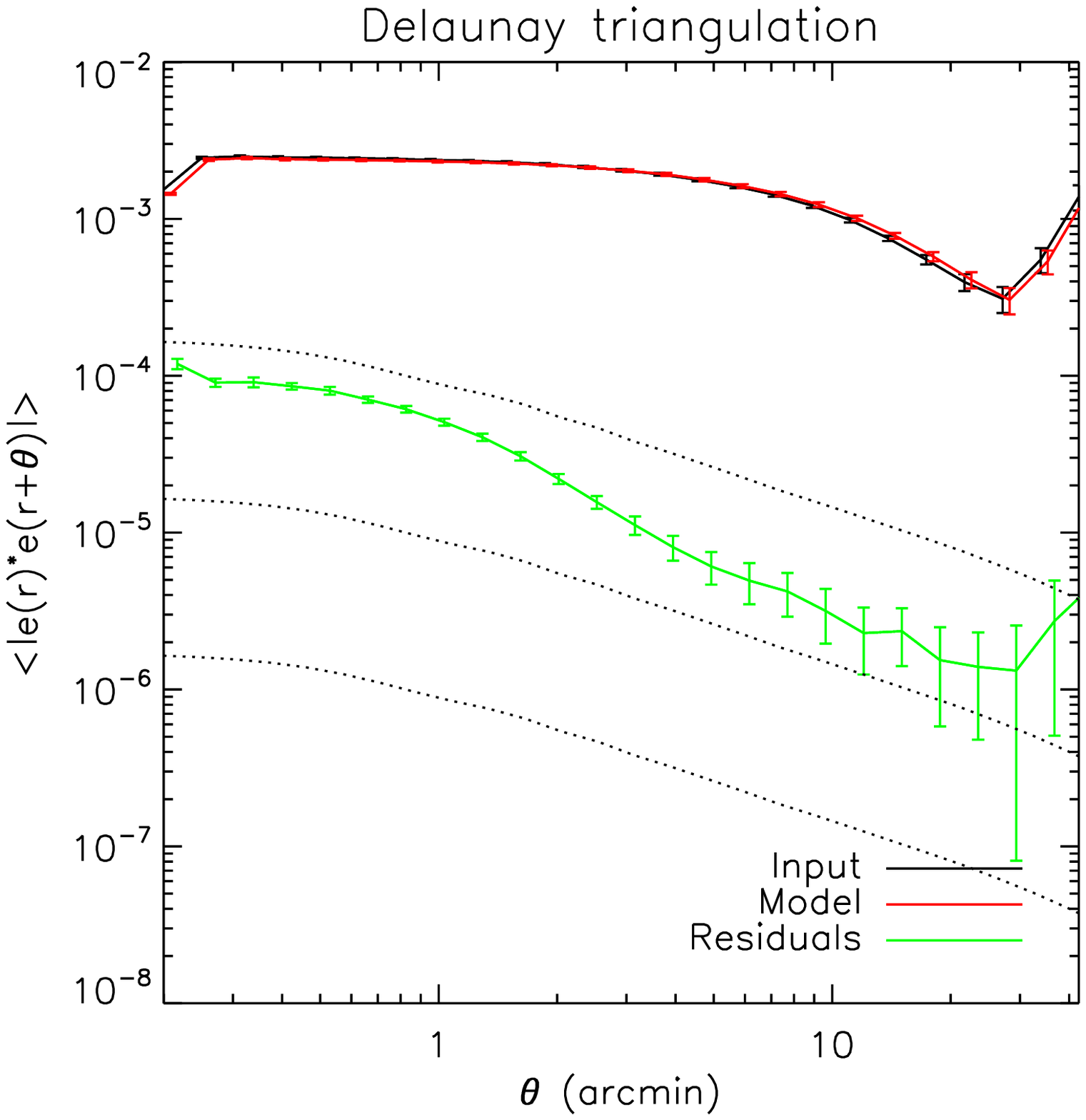}
\includegraphics[width=6cm,angle=0]{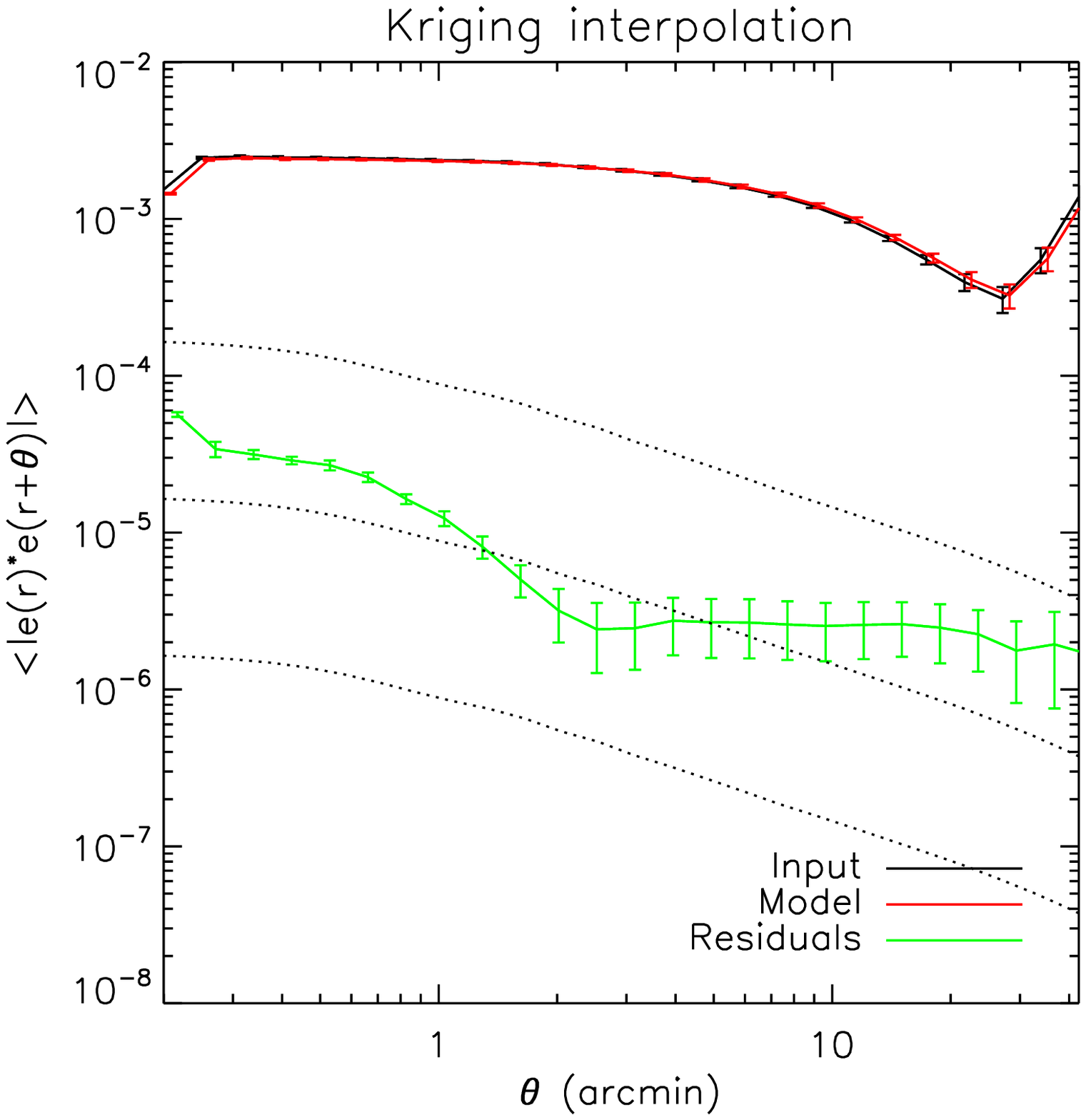}
\caption{PSF ellipticity correlation functions, averaged over 50 simulations. Left: polynomial interpolation. Center: Delaunay triangulation interpolation. Right: Kriging interpolation. In each panel, the black line represents the ellipticity correlation function of the input, the red line shows that of the model, and the green one shows that of the residuals. Dotted lines show the upper limits on the residuals that one must satisfy so that the weak lensing systematics $\sigma_{\rm sys}^2$ from the PSF are less than $10^{-5}$, $10^{-6}$ and $10^{-7}$ from top to bottom.} \label{fig_cor}
\end{figure*}

In Fig. \ref{fig_ecomp}, we compare the distribution of the ellipticity residuals, between the three considered interpolation techniques, for 50 simulations. Figure \ref{fig_coeffscomp} provides similar comparisons for the first four Principal Components of the PSF shape.
These figures, showing that the distribution of residuals is clearly tightest for the Kriging interpolation, confirm our claims above about how much better this technique performs. Furthermore, no significant correlation is visible between the residuals obtained with a polynomial interpolation and the other two interpolation schemes; they perform well even where a polynomial interpolation performs badly. On the other hand, the residuals obtained with a Kriging and those obtained with a Delaunay triangulation interpolations are correlated. This shows that since they are both sensitive to small scale information, their behaviors are comparable. For instance, they suffer similar limitations when the field varies quickly (as shown in Figs. \ref{fig_ell}-\ref{fig_coeff0}).

\subsection{Ellipticity correlation functions} \label{ssect_corfct}

We now turn to the PSF ellipticity correlation functions. If the interpolation is correct, the correlation function of the residuals should be consistent with zero, or at least negligible compared with that of the measured PSF ellipticity. In particular, \cite{rowe10} has shown that the correlation function of the residuals helps diagnose an under- or an over-fitting, and therefore is recognized a perfect test to choose between different interpolations. 

The ellipticity correlations functions are defined as $\chi(|\mathbf{\theta}|)=<e(\mathbf{r})e^*(\mathbf{r+\theta})>$, where $e$ is the complex ellipticity of either the input simulation, the interpolated PSF, or the residuals, and $e^*$ is the ellipticity's complex conjugate. In case of the residuals, the correlation function is identical to the function $D_1(\theta)$ defined by \cite{rowe10} ($D_1(\theta)\equiv<\delta e(\mathbf{r})^*\delta e(\mathbf{r}+\theta)>$, where $\delta e$ are the residuals). Although on real data, the PSF ellipticity correlation functions can only be estimated with stars, here we measure them using only the `galaxy-PSFs' in our simulations, that is, the measurement of the PSF on points not used to perform the interpolation. Hence, we discard the stars, which prevents our estimation of the residuals' correlation function from being underestimated due to a perfect interpolation on stars. Our measurements of the correlation functions therefore only probe the reconstruction of the PSF where it is not known a priori (and used for the interpolation).

Figure \ref{fig_cor} shows the PSF ellipticity correlation functions, averaged over 50 simulations, for a bivariate-polynomial, a Delaunay triangulation, and a Kriging interpolations from left to right. The black solid lines represent the ellipticity correlation functions as measured on the simulations. The red lines are those of the interpolated PSF. The green lines are those of the residuals. 
The correlation function of the residuals in Fig. \ref{fig_cor} confirm that a Kriging interpolation yields significantly better results than a polynomial interpolation. The difference is less striking with a Delaunay triangulation, although clearly noticeable. We discuss the effects on systematics in the context of weak gravitational lensing in section \ref{ssect_wlsys}.

\section{Discussion} \label{sect_discussion}

In this section, after discussing the level of systematics from the PSF interpolation for weak lensing, we discuss the stability of the techniques presented above. In particular, we focus on their dependence on the stellar S/N and stellar density. 

\subsection{Weak gravitational lensing systematics} \label{ssect_wlsys}

Although considered a premier cosmological probe, weak lensing is such a small effect that all possible systematic effects must be tackled very thoroughly. The PSF is the most significant systematic, making the correction of shapes for the PSF effects of paramount importance. Here, we quantify the systematics on the cosmic shear power spectrum due to PSF interpolation errors using the three schemes mentioned above. 

We assume that the PSF deconvolution, which occurs after its interpolation when measuring galaxy shapes, does not bring additional systematics, so that all systematics attributable to the PSF come from its modeling only. 
We compute the impact on the shear correlation function by following the definition by \cite{rowe10} and \cite{sph09}:
\begin{equation} \label{eq_dxi}
|\delta\xi_+^\gamma| (\theta) \leqslant \left| \frac{D_1(\theta)}{(P^\gamma)^2} \left< \left( \frac{R_{\rm PSF}}{R_{\rm gal}}\right)^4 \right> \right|,
\end{equation}
where $D_1$ is the correlation function of the residuals (see above), $P^\gamma$ is the shear susceptibility, and $R_{\rm PSF}$ and $R_{\rm gal}$ are the sizes of the PSF and of galaxies. We take the same values as those used by \cite{rowe10}: $P^\gamma=1.84$, $<(R_{\rm PSF}/R_{\rm gal})^4> \approx (1/1.5)^4$. Note that in Eq. (\ref{eq_dxi}), we ignore the second term of \cite{sph09}'s Eq. (15), which depends on the amplitude of the pre-correction PSF ellipticity. Indeed, since it puts the emphasis on the pre-correction PSF, this strong dependence makes it less relevant to our discussion.

The impact on the shear correlation function $\delta\xi_+^\gamma$ can be integrated in Fourier space, to estimate the level of systematics, in the sense of \cite{amara08}, coming from the imperfect PSF interpolation:
\begin{equation}
\sigma_{\rm sys}^2=\frac{1}{2\pi}\int |C_l^{\rm sys}|\ell (\ell+1) {\rm d}\ln \ell.
\end{equation}

The dotted lines in Fig. \ref{fig_cor} show the upper limits on the residuals ellipticity correlation function (assumed to be a constant fraction of the shear correlation function on all scales) such that the systematics due to the imperfect PSF modeling are less than $\sigma_{\rm sys}^2=10^{-5}, 10^{-6}, 10^{-7}$ from top to bottom. We use a $\Lambda$CDM cosmology with $\sigma_8=0.8$ to estimate those limits. Due to our neglecting the dependence of $\sigma_{\rm sys}^2$ on the pre-correction PSF ellipticity, the limits plotted in Fig. \ref{fig_cor} are optimistic compared to those we could achieve in a real survey, but give a good general sense of the required level of the PSF residuals.

\cite{amara08} investigated the allowed values of $\sigma_{\rm sys}^2$ such that, for a given survey, errors in the estimation of cosmological parameters are dominated by statistical errors. Table \ref{tab_surveys} gives the maximum $\sigma_{\rm sys}^2$ allowed for some current and planned surveys, computed using \cite{amara08}'s scaling relation (21).
By comparing this table with Fig. \ref{fig_cor}, while it appears that a simple bivariate polynomial interpolation does not allow one to reach the limit on the PSF systematics on current surveys such as the CFHTLS Wide, a Kriging approach lowers the residuals enough to meet the requirements. However, the techniques explored in this paper would not allow us to lower the systematics to the level needed for future ambitious wide field surveys.

Nevertheless, our analysis is made from mock ground-based images, and our techniques are restricted to single-field interpolation. 
More complex PSF interpolation schemes can be thought of. For instance, a multi-field interpolation scheme would allow one to look for coherent pattern from image to image, and therefore improve on the PSF model by taking into account more information than available in a single image. Examples of such techniques are provided by \cite{jarvis04} and \cite{jee11}; a similar approach has been undertaken in COSMOS by \cite{schrabback10}.
To sum up, our conclusions stand for PSF modeling based on single-field interpolation of the PSF from ground-based imaging surveys. More elaborate, multi-field interpolation techniques, either on ground-based or on space-based data, should allow one to lower the level of residuals, and reach the requirements for Stage IV weak lensing surveys.

\begin{table}
\caption{Maximum level of systematics on shear power spectrum for weak lensing surveys}
\begin{center}
\begin{tabular}{|c|c|c|c|c|}
\hline
Survey & Area & Galaxy density  & Median & $\sigma_{\rm sys}^2$ \\
& (deg$^2$) & (arcmin$^{-2}$) & redshift \\
\hline
CFHTLS & 170 & 12 & 0.7 & $2\times10^{-6}$ \\
COSMOS & 2 & 50 & 0.9 & $8\times10^{-6}$ \\
Stage IV & 15000 & 30 & 0.9 &  $1.2\times10^{-7}$\\
\hline
\end{tabular}
\end{center}
\label{tab_surveys}
\end{table}


\subsection{Impact of stars' S/N -- shape measurement errors vs interpolation errors} \label{ssect_sn}

\begin{figure}
\centering
\includegraphics[width=8cm,angle=0]{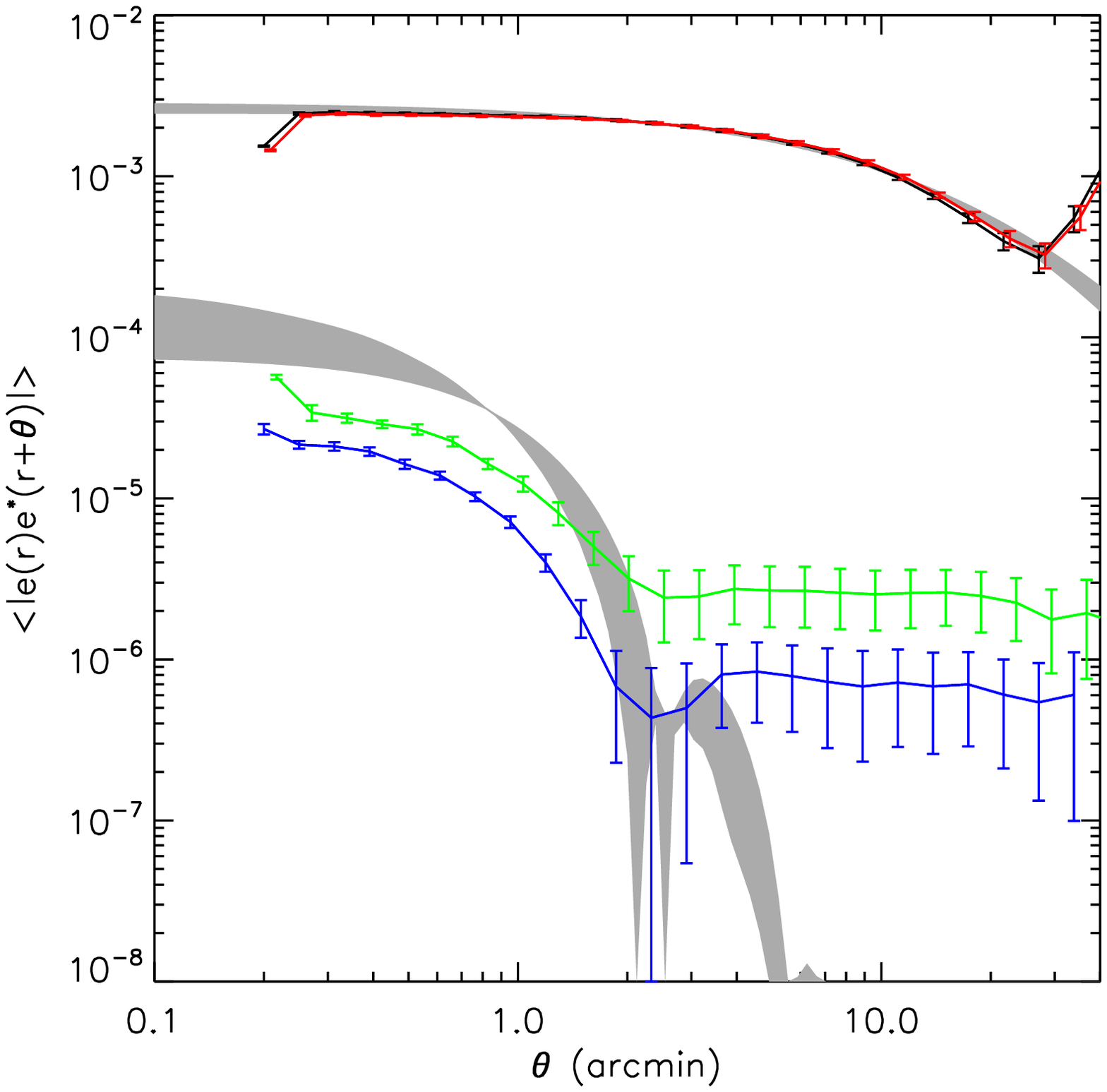} 
\caption{Ellipticity correlation functions, with a Kriging interpolation, for different S/N stars. The black and red lines are the input and model correlation functions for S/N=100 stars. The green line is the correlation function of the residuals for S/N=100 stars. The blue line is the correlation function of the residuals for S/N=700 stars. The shaded region in the upper part is the region spanned by the correlation functions of theoretical Gaussian random fields used to estimate the optimal correlation function of residuals after Wiener-filtering (lower shaded region).} \label{fig_corhsn}
\end{figure}

Although so far we have only focussed on the interpolation of the PCA coefficients, our analysis includes the first step of a PSF modeling process, namely the PSF's shape measurement (PCA decomposition in our pipeline). 

The non-vanishing ellipticity correlation functions of the residuals (Fig. \ref{fig_cor}) betray a dependence of the PSF model on the stellar S/N, defined as the ratio of the mean of the star's pixels to the r.m.s of the background. 
To test this effect of the S/N, we create a similar set of simulations, with higher S/N stars (S/N=700 for each star). 
The blue line in Fig. \ref{fig_corhsn} shows the correlation function of the residuals for a Kriging interpolation with those higher S/N stars, and compares it with the residuals obtained previously with stars of S/N=100 (green line). It is obvious that the residuals have been shifted downwards, due to the better shape measurement in the first step of our pipeline.

To discriminate the errors from the shape measurements against those from the interpolation, we estimate the minimal errors on the PSF modeling that we can expect in the ideal case of a Wiener-filtered Gaussian random field (\citealt{amara10}). In this case, the model is optimal (i.e., it gives the lowest achievable residuals), and depends only on the number of stars and on their S/N.
The shaded region in the upper part of Fig. \ref{fig_corhsn} shows the span in ellipticity correlation functions of 12 Gaussian random fields with power spectrum resembling that of our simulations. The shaded region in the lower part of the figure is the span of the corresponding correlation functions of the residuals, for stars with S/N=100. At small scales, the correlation function of the residuals reflects the errors from the shape measurement. For increasing scales, the interpolation cancels out those errors, and the residuals tend to zero at large scales.
Although these perfect theoretical expectations cannot be rigorously compared with our simulations, since they are made in a different regime (Gaussian vs non-Gaussian fields interpolated with different techniques), they give a good indication about how the errors from the interpolation compare with those from the shape measurement in our analysis. At small scales, the order of magnitude of the correlation function of the residuals (green line) being similar to that of the theoretical estimation, we can safely conclude that our Kriging interpolation is likely close to optimal. 
Our theoretical estimates do not allow us to explain the presence of a plateau in the correlation function of the residuals at large scale, since at large scales, Kriging is underperforming compared with theoretical expectations. But the increase of the pre-correction ellipticity correlation function at large scales in our simulations, which is not taken into account in our theoretical estimations, may explain the presence of these plateaus.

Finally, we tested the robustness of the interpolation techniques against the stellar S/N (and hence, the reliability of the stars' model) by creating another set of simulations, with a realistic S/N distribution, ranging from S/N=20 to S/N=1000, and peaking around S/N=100. The total number of stars is unchanged. Lowering the S/N of some stars decreases the fidelity of the associated PCA model, and potentially perturbs the interpolation, since the coefficients on which it is based are more noisy. We find that ignoring low S/N stars does not impact the interpolation, as long as high enough S/N ($\geqslant 100$) stars are sufficiently numerous. This result was already mentioned by \cite{sph09} as part of their work on PSF shape measurement. They claimed that $\approx 1$ star per square arcminute with S/N$\geqslant$100 was necessary to obtain a reliable model.

\subsection{Global vs local stellar density} \label{ssect_density}

Here, we investigate the dependence of the interpolation techniques on the number (identically, the density) of stars. We compare the fidelity of the interpolation at a given position, as a function of the local stellar density and as a function of the global stellar density in the image. In the former case, we naively expect interpolations to be better constrained in high density regions than in lower density regions. In the latter case, we expect interpolations to perform better with more stars.

\begin{figure}
\centering
\includegraphics[width=7cm,angle=0]{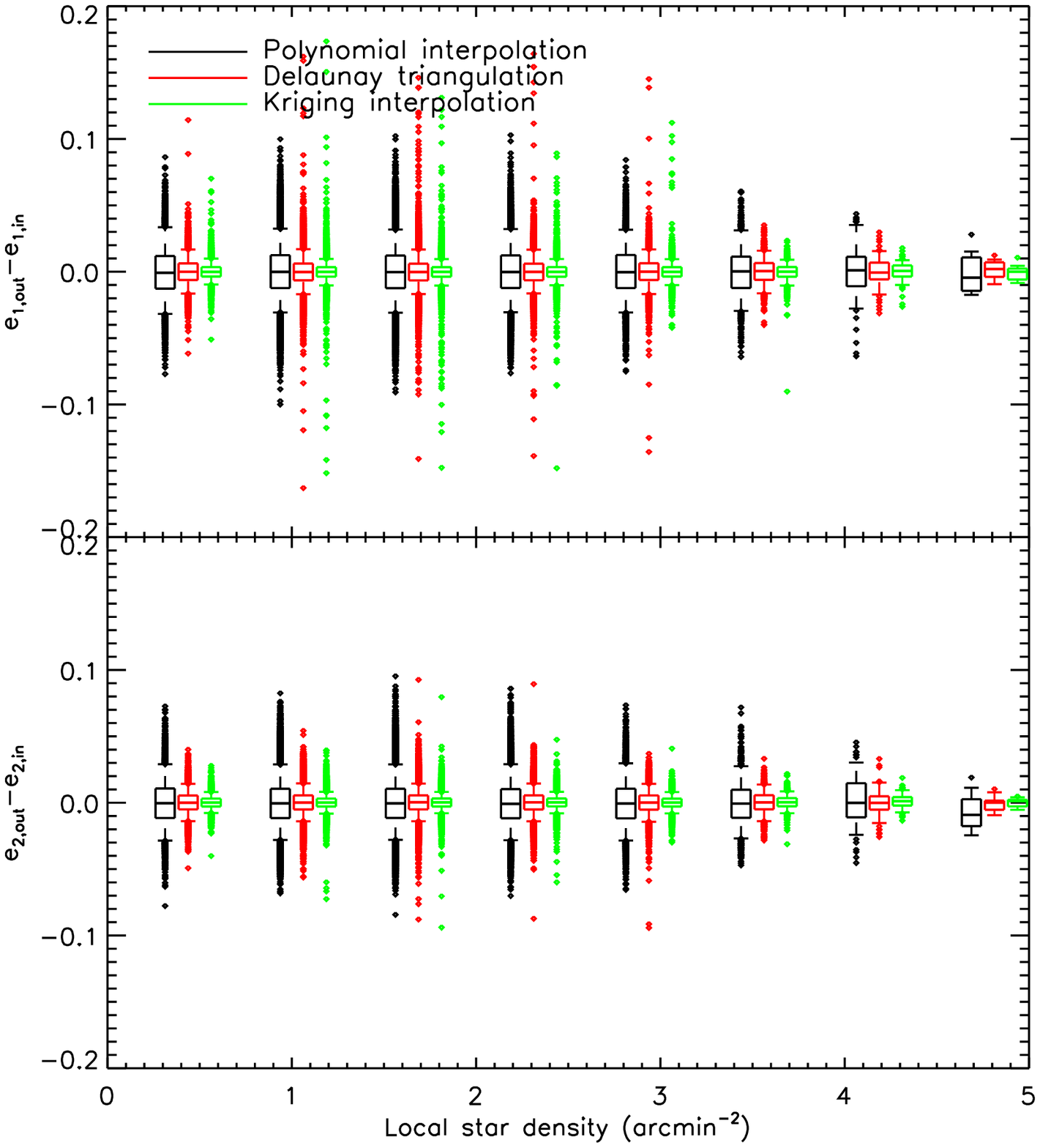} 
\caption{Ellipticity residuals box plots as a function of local stellar density, for polynomial interpolation (black), Delaunay triangulation (red) and Kriging (green). For each box-plot, the central line represents the median of the distribution of residuals, the adjacent boxes are the central quartiles, and the symbols are the 5\% outer points. For each local stellar density, box-plots are offsets for visibility.} \label{fig_densell}
\end{figure}

\begin{figure}
\centering
\includegraphics[width=7cm,angle=0]{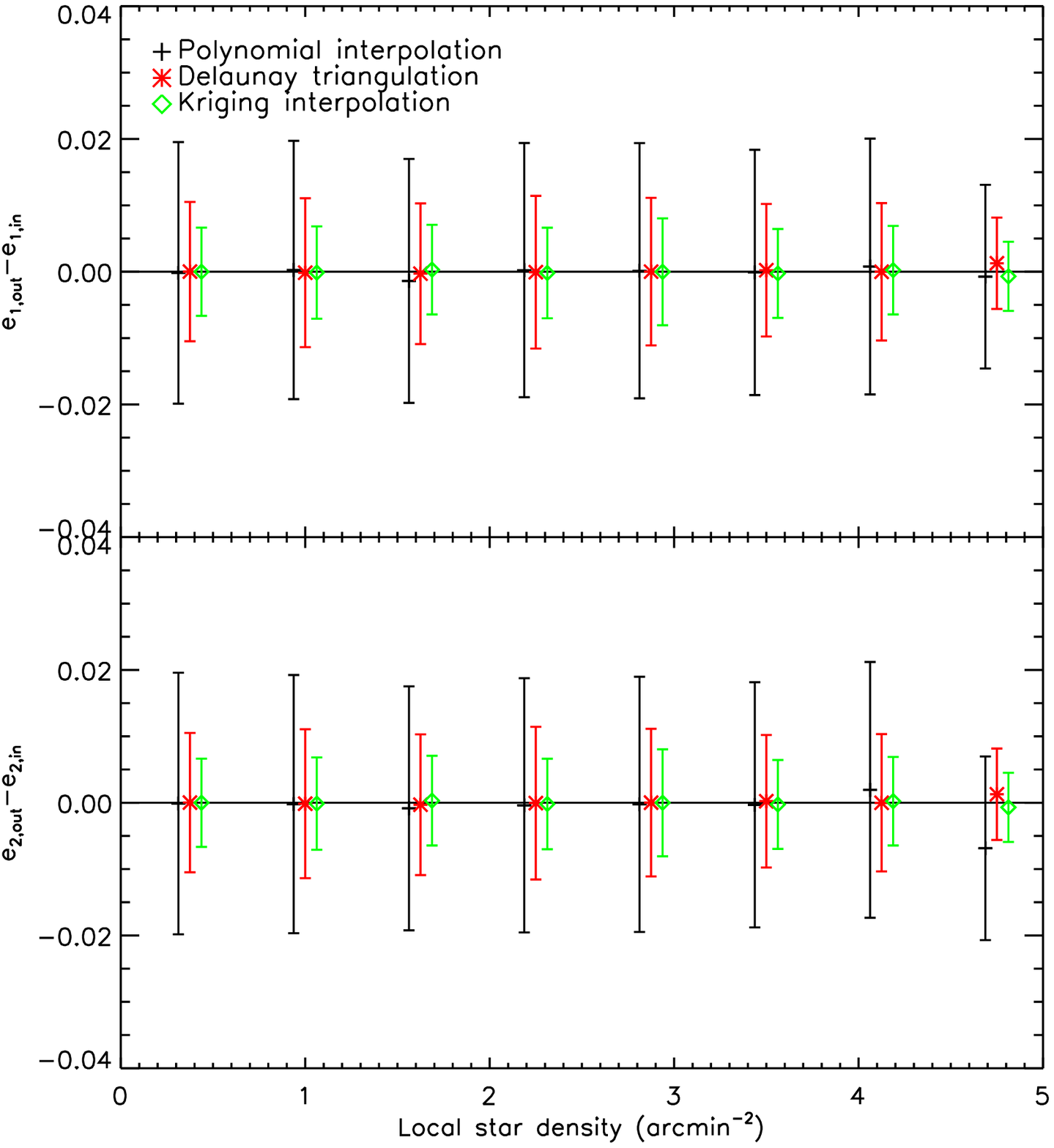} 
\caption{Median and rms of ellipticity residuals distributions as a function of stellar density. The color code is the same as in Fig. \ref{fig_densell}.} \label{fig_densellvar}
\end{figure}

Figure \ref{fig_densell} shows box-plots of the distribution of the residuals of the two ellipticity components, as a function of the local star density, computed in one-arcmin-radius circular apertures around `galaxy-PSFs', for a polynomial interpolation (black), a Delaunay triangulation (red) and a Kriging interpolation (green). Each distribution is drawn from the set of 50 simulations used in section \ref{sect_results}. For each box-plot, the central line represents the median of the distribution, the boxes around it are the second and third quartiles, and the external symbols are the outer 5\% points. Figure \ref{fig_densellvar} shows the r.m.s of those distributions, as a function of local stellar density.

These figures concur with our claims above, that the Kriging interpolation gives the smallest residuals. This is the case for all stellar densities. They also show that the variance of the residuals, as well as central quartiles of their distributions, does not exhibit a notable dependence on the local stellar density. Therefore, high density regions do not constrain the interpolation more than lower density regions. However, it is clear from Fig. \ref{fig_densell} that the width of the full distribution of residuals decreases in high stellar density regions: outliers from interpolation failures are much less likely in high stellar density region than in low density regions. This observation can provide a way to weigh the PSF model in an image.

\begin{figure}
\centering
\includegraphics[width=7cm,angle=0]{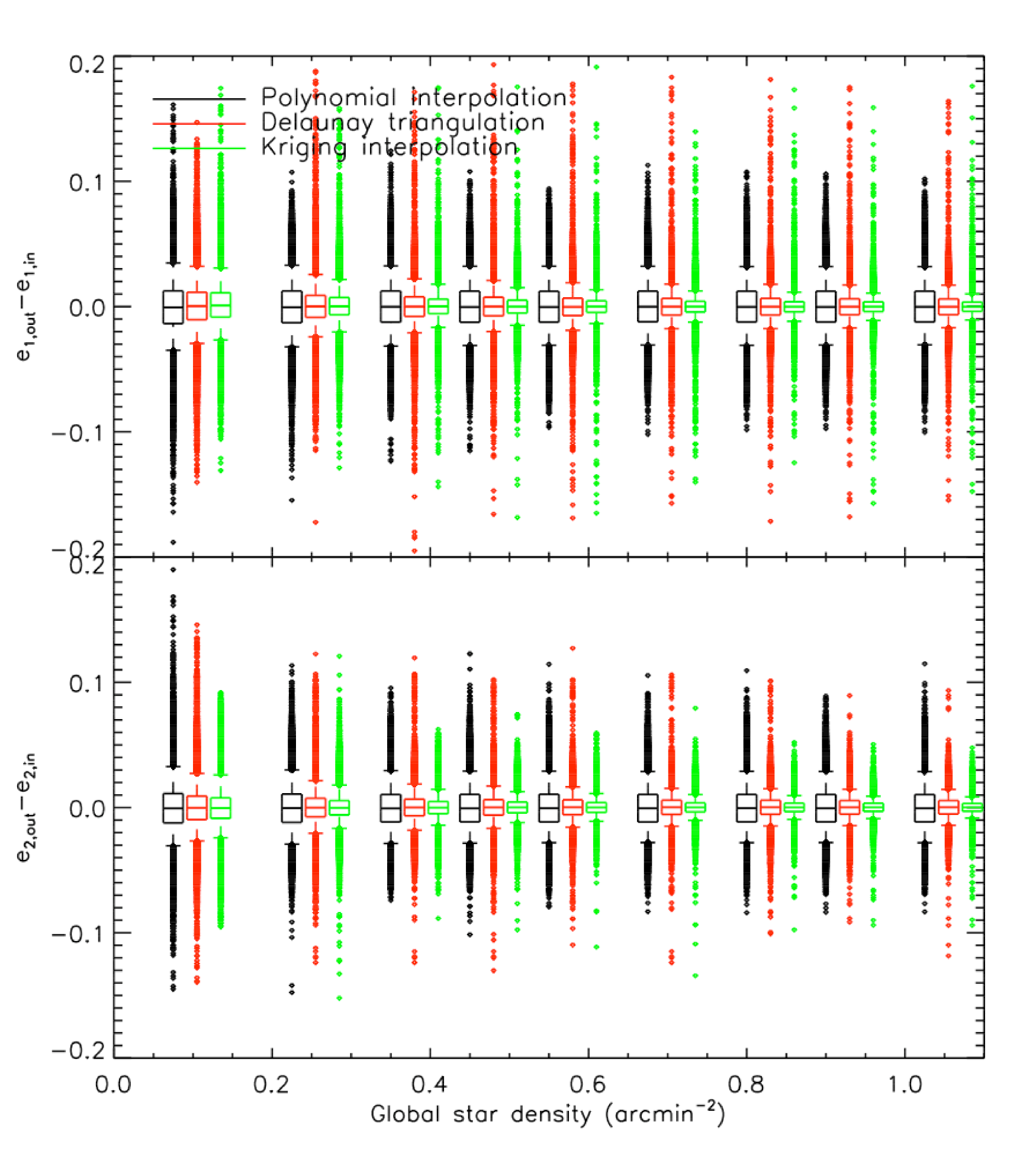} 
\caption{Same as Fig. \ref{fig_densell}, but for global star density.} \label{fig_densellglob}
\end{figure}

Figure \ref{fig_densellglob} shows a similar box-plot as Fig. \ref{fig_densell}, but for ellipticity residuals as a function of the global stellar density (i.e., as a function of the total number of stars available in the image to perform the interpolation). The accuracy of a polynomial interpolation does not show a strong dependence on the total number of star. This is because, for a low enough order polynomial, the general pattern of the polynomial is set by a small number of stars; adding stars does not help constrain the polynomial better. On the opposite, a Delaunay triangulation and Kriging use all information at small scales: their dependence on the total number of stars is significant.

To sum up this section, the interpolation accuracy mostly depends on the global star density rather than on the local star density, especially for Delaunay triangulation and Kriging. While the global star density acts on the general accuracy of the interpolation, the local star density helps eliminate outliers from the interpolation.


\subsection{Sensitivity to outliers}

Each star in an image provides an imperfect realization of the PSF, which is pixelated and noisy. Furthermore, the PSF shape measurement does not give a perfect description of each star. From those facts, outliers are possible, which could be seen e.g. as stars with unphysical ellipticity.  Such outliers can be source of error in the interpolation. For instance, one can see a small outlier in the stars' ellipticity on Fig. \ref{fig_ell}, at position $(x,y)\approx(1500,5500)$. It is visible that the Delaunay triangulation and the Kriging interpolation are affected by this outlier, since their residuals are clearly bigger than average around this position.

We find that, due to their being local interpolation schemes, the Delaunay triangulation and the Kriging interpolation are more prone to outliers than a polynomial interpolation, which tends to smooth the spatial pattern out (however, a very high order polynomial interpolation would also become more prone to outliers). 

This sensitivity to outliers is alleviated by a simple control of outliers. We perform an Interquartile range test on the distribution of each Principal Component. 
For each component, outliers are defined as having values less than $Q_1-\alpha {\rm IQR}$ or more than $Q_3+\alpha {\rm IQR}$, where $Q_i$ is the $i$th quartile of the component's distribution, ${\rm IQR}=Q_3-Q_1$ is the interquartile range, and $\alpha$ is a free parameter.
As soon as at least one component of a star is found to be an outlier, this star is discarded. Using a loose criterium ($\alpha=3.5$) helps control the sensitivity to outliers, while removing only a negligible proportion of stars (typically 1\%).

\section{Conclusion} \label{conclusion}

We investigated how to improve on the techniques usually used to interpolate the PSF. An extremely reliable PSF model is necessary, in particular for weak gravitational lensing analyses, to beat systematics down. Any PSF model relies on a robust interpolation of any quantity used to describe the PSF (e.g. its moments, or the coefficients of a decomposition on a given set of basis functions).

To this end, we developed simulations of PSF fields. The simulations are based on a typical Subaru SuprimeCam image, whose stars are decomposed into Principal Components, which are then used as basis functions to build mock stars. The background of the simulations is given the same statistics as that of the real image used as a reference for our simulations.

We used the simulations to compare several interpolations schemes, including bivariate polynomial, radial basis functions, Delaunay triangulation and Kriging. 
We assumed that the first step of PSF modeling (its shape measurement --for instance, its decomposition into PCA in our analysis) is well constrained and brings only negligible errors on the model. Therefore, we focussed on the interpolation part of the pipeline, that is, the interpolation of the principal components coefficients of the stars.
We found RBF to be unstable. We found a Kriging approach to bring the best results, slightly better than a Delaunay triangulation, and significantly better than the traditionally used bivariate polynomial interpolation.

We showed how the full PSF modeling depends upon the S/N of stars, and how the Kriging interpolation is close to optimal. We discussed how the different interpolation schemes depend on the local and global star densities: models are better controlled in high stellar density regions; Delaunay triangulation and Kriging are more sensitive than a polynomial interpolation to the total number of stars. We noticed that Kriging and Delaunay interpolations are more prone to outliers than a polynomial interpolation; however, a simple control on outliers helps cancel this difficulty. Finally, although Kriging is numerically more expensive than a polynomial interpolation, it is worth the improvement it provides.

The ellipticity correlation functions of the residuals between the input PSF and the interpolated PSF allowed us to conclude that although a Kriging interpolation is sufficiently accurate for current weak lensing surveys, it will not allow us, as used in this paper, to meet the requirement on the control of systematics for upcoming ambitious surveys. Since Kriging is considered as the best linear unbiased interpolator, it seems difficult to go beyond it to make significant improvements. However, our analysis stands for  single-field interpolation, i.e., the PSF is interpolated on each image separately, with a maximum density of useful stars about one per square arcminute. More elaborate techniques, relying on several images to perform the interpolation, will improve on the results of this paper. For example, when the PSF is know to be stable in time (i.e., for space-based observations), stacking several images allows one to artificially increase the density of stars, and therefore to better constrain the interpolation. Another possibility is to look for coherent patterns in different images, even for ground-based data, which allows for a multi-field interpolation to be performed. Those new techniques will surely permit to meet the requirements for future surveys.

\section*{Acknowledgments}

We want to thank Barney Rowe and Alexandre R\'efr\'egier for useful discussions. We thank Satoshi Miyazaki for providing us with Subaru weak lensing images. We also thank Richard Massey for his comments on the manuscript, as well as the anonymous referee for their useful comments.
JB acknowledges support from HST grant AR-11747.
SP acknowledges support from the Caltech Summer Undergraduate Research Fellowship (SURF) program, through the Elachi endowment and internal JPL research funding.
Part of this work was carried out at Jet Propulsion Laboratory, California Institute of Technology, under a contract with NASA.

\label{lastpage}

\end{document}